\documentclass[preprint, 12pt]{elsarticle}

\biboptions{sort&compress}

\usepackage{amssymb}
\usepackage{latexsym}

\usepackage{url}
\usepackage{xcolor}
\definecolor{newcolor}{rgb}{.8,.349,.1}

\journal{Journal of Computational Physics}

\usepackage[T1]{fontenc}
\usepackage[utf8]{inputenc}
\usepackage{textcomp}
\usepackage{amsmath,amsfonts,braket,mathtools,graphicx,amsthm}
\usepackage[font=footnotesize]{caption}
\usepackage{bm} 
\usepackage{blindtext}
\usepackage{hyperref}
\usepackage{multirow} 

\usepackage{tikz}
\usetikzlibrary{shapes.geometric, arrows, arrows.meta}

\tikzstyle{arrow} = [thick,->,>=stealth]
\usepackage{varwidth}
\usepackage{colortbl}

\usepackage[plain]{algorithm}
\usepackage{algpascal}

\usepackage{algpseudocode}
\algnewcommand\algorithmicinput{\textbf{Input:}}
\algnewcommand\Input{\item[\algorithmicinput]}

\algnewcommand\algorithmicoutput{\textbf{Output:}}
\algnewcommand\Output{\item[\algorithmicoutput]}

\algnewcommand\algorithmicinitialize{\textbf{Initialize:}}
\algnewcommand\Initialize{\item[\algorithmicinitialize]}

\newcommand{\bmtheta}{\bm{\theta}}
\newcommand{\bmp}{\bm{p}}
\newcommand{\bmz}{\bm{z}}

\DeclarePairedDelimiter\abs{\lvert}{\rvert}

\DeclareMathOperator*{\argmin}{arg\,min}
\makeatletter
\let\oldabs\abs
\def\abs{\@ifstar{\oldabs}{\oldabs*}}

\usepackage{caption, subcaption}
\DeclareCaptionLabelSeparator{AIAalgorithm}{--}

\usepackage{etoolbox}

\makeatletter
\def\ps@pprintTitle{%
 \let\@oddhead\@empty
 \let\@evenhead\@empty
 \def\@oddfoot{}%
 \let\@evenfoot\@oddfoot}
\makeatother

\begin{document}

\begin{frontmatter}

\title{Adaptive multi-stage integration schemes for Hamiltonian Monte Carlo}

\author[1]{Lorenzo Nagar} \corref{cor1}
\ead{lnagar@bcamath.org}
\cortext[cor1]{Corresponding author:}
\author[2]{Mario Fernández-Pendás}
\author[3]{Jesús María Sanz-Serna}
\author[1,4]{Elena Akhmatskaya}

\address[1]{BCAM - Basque Center for Applied Mathematics, Alameda de Mazarredo 14, 48009 Bilbao, Spain}
\address[2]{DIPC, Donostia International Physics Center, Manuel Lardizabal Ibilbidea 4, 20018 Donostia, Spain}
\address[3]{Departamento de Matemáticas, Universidad Carlos III de Madrid, Avenida Universidad 30, 28911 Leganés, Spain}
\address[4]{Ikerbasque - Basque Foundation for Science, Euskadi Plaza 5, 48009 Bilbao, Spain}

\begin{abstract}
Hamiltonian Monte Carlo (HMC) is a powerful tool for Bayesian statistical inference due to its potential to rapidly explore high dimensional state space, avoiding the random walk behavior typical of many Markov Chain Monte Carlo samplers. The proper choice of the integrator of the Hamiltonian dynamics is key to the efficiency of HMC. It is becoming increasingly clear that multi-stage splitting integrators are a good alternative to the Verlet method, traditionally used in HMC. Here we propose a principled way of finding optimal, problem-specific integration schemes (in terms of the best conservation of energy for harmonic forces/Gaussian targets) within the families of 2- and 3-stage splitting integrators. The method, which we call Adaptive Integration Approach for statistics, or s-AIA, uses a multivariate Gaussian model and simulation data obtained at the HMC burn-in stage to identify a system-specific dimensional stability interval and assigns the most appropriate 2-/3-stage integrator for any user-chosen simulation step size within that interval. s-AIA has been implemented in the in-house software package \textsf{HaiCS} without introducing computational overheads in the simulations. The efficiency of the s-AIA integrators and their impact on the HMC accuracy, sampling performance and convergence are discussed in comparison with known fixed-parameter multi-stage splitting integrators (including Verlet). Numerical experiments on well-known statistical models show that the adaptive schemes reach the best possible performance within the family of 2-, 3-stage splitting schemes.
\end{abstract}

\end{frontmatter}

\section{Introduction}\label{sec:Introduction}
\begin{sloppypar}
First introduced for lattice field theory simulations \cite{duane1987hmc}, Hamiltonian Monte Carlo (HMC) is nowadays recognized as a popular and efficient tool for applications in Bayesian statistical inference \cite{neal2011mcmc}.
\end{sloppypar}

Using gradient information on the posterior distribution, HMC reduces random walk behavior typical of many conventional Markov Chain Monte Carlo (MCMC) samplers and makes it possible to sample high dimensional and complex distributions more efficiently than simpler MCMC algorithms. The use of Hamiltonian dynamics makes HMC able to perform large moves while keeping high acceptance rates, thus  lowering the correlation between samples, provided that an accurate symplectic integrator is in use \cite{numerical_hamiltonian_problems, bou-rabee_sanz-serna_2018}. On the other hand, known drawbacks of HMC are the computational cost deriving from the evaluation of gradients and the strong dependence of the performance on the choice of the parameters in the algorithm. Many variants of HMC have been proposed in the literature during the last decades (see \cite{radivojevic_akhmatskaya_MHMC_2020} for an advanced list of HMC methods in computational statistics and physical sciences).

Numerical integration of the Hamiltonian equations of motion is crucial for HMC, since its accuracy and efficiency strongly affect the overall performance of the method. Velocity Verlet \cite{verlet1967, swope1982} is currently the method of choice owing to its simplicity, optimal stability properties and computational efficiency. Recently proposed multi-stage splitting integrators have shown promising performance in HMC for statistical and molecular simulation applications \cite{bcss2014, campos_sanz-serna2017, calvo2021hmc}. Such integrators are as easy to implement as Verlet schemes due to their kick-drift structure. However, they possess shorter \emph{stability intervals}\footnote{Stability interval is defined as the largest interval of step sizes for which the integrator stays stable, i.e. the numerical solution remains bounded as the number of computed points increases when the integrator is applied to the harmonic oscillator \cite{bcss2014}.} than corresponding multi-stage Verlet algorithms \cite{bcss2014}.

The Adaptive Integration Approach (AIA) \cite{AIApaper2016} for HMC and its extensions MAIA and e-MAIA for Modified HMC (MHMC) methods \cite{MAIApaper2017} offer an intelligent (system- and step size-specific) choice of the most appropriate 2-stage integrator in terms of the best conservation of energy for harmonic forces. They have been formulated and implemented for molecular simulation applications and demonstrated an improvement in accuracy, stability and sampling efficiency compared with the fixed-parameter 1-, 2-stage numerical integrators (including the standard Verlet) when used in simulations of complex physical systems \cite{MAIApaper2017, AIApaper2016, bonillaetal2021, bonilla2021batteries, bonilla2022interfacial, escribano2017}.

In this paper, we propose an Adaptive Integration Approach for statistics, that we call s-AIA, which extends the ideas of the original AIA to Bayesian statistical inference applications. The method employs a theoretical analysis of the multivariate Gaussian model and simulation data obtained at the HMC burn-in stage to identify a system-specific dimensional stability interval and assigns the most appropriate 2-, 3-stage integrator at any user-chosen simulation step size within that interval. To construct s-AIA, we address the difficulties encountered by the extension to  the computational statistics scenario of the assumptions typical of molecular simulation applications made in AIA --- such as dominating harmonic forces, known angular frequencies and resonance conditions, nonrandomized integration step size. The proposed algorithm does not add computational overheads during a simulation.

We have implemented s-AIA in the in-house software \textsf{HaiCS} (Hamiltonians in Computational Statistics) \cite{tijana_thesis, radivojevic_akhmatskaya_MHMC_2020} and tested its efficiency and impact on the HMC accuracy, sampling performance and convergence in comparison with known fixed-parameter multi-stage splitting integrators for HMC-based methods (including Velocity Verlet). The numerical experiments have been performed on representative benchmarks and datasets of popular statistical models.

The paper is structured as follows. We briefly review HMC in Section~\ref{sec:HMC} and multi-stage integrators in Section~\ref{sec:MultiStageIntegrators}. The s-AIA algorithm and its implementation are presented in Section~\ref{sec:sAIA}. Validation and testing of the new algorithm are described and discussed in Section~\ref{sec:NumericalResults}. Our conclusions are summarized in Section~\ref{sec:Conclusion}.

\section{Hamiltonian Monte Carlo}\label{sec:HMC}
\begin{sloppypar}
Hamiltonian Monte Carlo (HMC) is a Markov Chain Monte Carlo (MCMC) method for obtaining correlated samples $\bm{\theta}_i \sim \pi(\bm{\theta})$ from a target probability distribution $\pi (\bmtheta)$ in $\mathbb{R}^D$ by generating a Markov chain in the joint phase space $\mathbb{R}^D \times \mathbb{R}^D$ with invariant distribution
\end{sloppypar}
\begin{equation}\label{eq:InvariantJointDistribution}
\pi(\bm{\theta}, \bm{p}) = \pi(\bm{\theta}) p(\bm{p}) \propto \exp (- H(\bm{\theta}, \bm{p})).
\end{equation}
Here
\begin{equation}\label{eq:HamiltonianSeparable}
H(\bmtheta, \bmp) = K(\bmp)+U(\bmtheta) = \frac{1}{2} \bmp^T M^{-1} \bmp + U(\bmtheta)
\end{equation}
is the Hamiltonian function, where the potential energy $U(\bmtheta)$ is related to the target $\pi(\bm{\theta})$ by means of
\begin{equation*}
U(\bmtheta) = - \log \pi(\bmtheta) + \text{const} \, ,
\end{equation*}
and the kinetic energy $K(\bmp)$ is specified through an auxiliary momentum variable $\bmp$ drawn from the normal distribution $\mathcal{N}(0, M)$, with $M$ being a symmetric positive definite matrix (the mass matrix).

HMC alternates  momentum update steps, where a sample of $\bmp$  is drawn from the distribution $\mathcal{N} (0, M)$, with
steps where both position $\bmtheta$ and momenta $\bmp$ are updated through the numerical integration of the Hamiltonian dynamics
\begin{equation}\label{eq:HamiltonianSystemSeparable}
\frac{d \bm{\theta}}{dt} = M^{-1} \bm{p}, \qquad \frac{d \bm{p}}{dt} = - \nabla_{\theta} U(\bm{\theta}).
\end{equation}
The latter is performed using an explicit symplectic and reversible integrator. If $\Psi_h$ is the map in phase space that advances the numerical solution over a step size of length $h$, symplecticness means
\cite{numerical_hamiltonian_problems}
$$
\quad \Psi'_h (\bmtheta, \bmp)^T J^{-1} \Psi'_h (\bmtheta, \bmp) = J^{-1}, \quad \forall (\bmtheta, \bmp) \in \Omega, \, \forall h >0,
$$
where $\Psi'_h$ is the Jacobian matrix of $\Psi_h$, $\Omega$ is an open set in phase space,
$$J = \begin{pmatrix}
0 & I \\
-I & 0
\end{pmatrix},$$ and
$I$ is the $D\times D$ unit matrix. Reversibility demands
$
\Psi_h \circ \mathcal{F} = \left( \Psi_h \circ \mathcal{F} \right)^{-1},
$
where $\mathcal{F} (\bmtheta, \bmp) = (\bmtheta, - \bmp)$ is the  \emph{momentum flip} map.
Symplecticness and reversibility ensure that $\pi(\bmtheta, \bmp)$ is an invariant measure for the Markov chain.
Given the state of the Markov chain $(\bmtheta_i, \bmp_i)$ at the beginning of the $i$-th iteration, a proposal $(\bmtheta', \bmp')$ is obtained by integrating the Hamiltonian equations of motion for $L$ steps using $\Psi_h$, i.e.
\begin{equation}\label{eq:Integration}
(\bmtheta', \bmp') = \underbrace{\Psi_h \circ ... \circ \Psi_h}_\text{$L$ times} (\bmtheta_i, \bmp_i).
\end{equation}
Due to  numerical integration errors, the Hamiltonian energy and thus the target density \eqref{eq:InvariantJointDistribution} are not exactly preserved. The invariance of the target density is ensured through a Metropolis test with acceptance probability
\begin{equation*}
\alpha = \min \{1, \exp (- \Delta H) \},
\end{equation*}
where
\begin{equation}\label{eq:EnergyError}
\Delta H = H (\bmtheta', \bmp') - H (\bmtheta_i, \bmp_i)
\end{equation}
is the energy error resulting from the numerical integration. In case of acceptance, $\bmtheta'$ is the starting point for the following iteration, i.e. $\bmtheta_{i+1} = \bmtheta'$, whereas in case of rejection, the initial proposal $\bmtheta_i$ is kept for the following iteration, i.e. $\bmtheta_{i+1} = \bmtheta_i$. In both cases, the momentum is discarded and a new momentum $\bmp_{i+1}$ is drawn from its Gaussian distribution.

\subsection{Splitting}
The integration of the Hamiltonian dynamics in HMC is always performed by resorting to the idea of splitting. The split systems
\begin{align*}
\text{(A)} & \quad \frac{d \bmtheta}{dt} = \nabla_p K(\bmp) = M^{-1} \bmp, &\qquad &\frac{d \bmp}{dt} = - \nabla_{\theta} K(\bmp) = 0, \\
\text{(B)} & \quad \frac{d \bmtheta}{dt} = \nabla_p U(\bmtheta) = 0, &\qquad &\frac{d \bmp}{dt} = - \nabla_{\theta} U(\bmtheta),
\end{align*}
 have solution flows $\varphi^A_t$ and $\varphi^B_t$ explicitly given by
\begin{equation}\label{eq:SeparateHamiltonianFlows}
\varphi^A_t (\bmtheta, \bmp) = (\bmtheta + t M^{-1} \bmp, \bmp), \qquad \varphi^B_t (\bmtheta, \bmp) = (\bmtheta, \bmp - t \nabla_{\theta} U(\bmtheta));
\end{equation}
 these flows are often called a position \emph{drift} and a momentum \emph{kick} respectively. The integration of the target dynamics \eqref{eq:HamiltonianSystemSeparable} is carried out by combining drifts and kicks. The best known algorithm is the  Velocity Verlet integrator \cite{verlet1967, swope1982}
\begin{align}
\bmp & \leftarrow \bmp - \frac{h}{2} \nabla_{\theta} U(\bmtheta), \nonumber \\
\bmtheta & \leftarrow \bmtheta + h M^{-1} \bmp, \nonumber \\
\bmp & \leftarrow \bmp - \frac{h}{2} \nabla_{\theta} U(\bmtheta). \label{eq:1sVV}
\end{align}
 With the notation in \eqref{eq:SeparateHamiltonianFlows}, the algorithm may be written as
\begin{equation}\label{eq:1sVVmap}
\Psi_h^{\text{VV}} = \varphi^B_{\frac{h}{2}} \circ \varphi^A_h \circ \varphi^B_{\frac{h}{2}}.
\end{equation}
As before, $h$ is the length of an integration step, i.e. step size. 
By switching the roles of $A$ and $B$ in \eqref{eq:1sVVmap} one obtains the Position Verlet algorithm \cite{tuckerman1992}, whose performance is often worse than that of the velocity scheme \cite{bou-rabee_sanz-serna_2018}.

More general splitting integration schemes \cite{blanes_casas_murua2008, bou-rabee_sanz-serna_2018} that alternate position drifts and momentum kicks will be reviewed in Section~\ref{sec:MultiStageIntegrators}.

\subsection{Advantages and limitations of HMC}
By suitably choosing the time span $Lh$ of the numerical integration (cf. \eqref{eq:Integration}), HMC offers the possibility of generating proposals that are  sufficiently far from the current state of the Markov chain. At the same time, for fixed $Lh$, one may always reduce $h$ and increase $L$ to achieve a more accurate numerical integration and therefore an arbitrarily high acceptance rate. Thus HMC is in principle able to generate samples with low correlation and to explore rapidly the state space, even if the dimensionality is high, avoiding in this way the random walk behavior of simpler MCMC algorithms. Unfortunately, it is well known that in practice the performance of HMC very much depends on the choice of the parameters $h$ and $L$.

Since most of the computational effort in HMC goes in the (often extremely costly) evaluations of the gradient $\nabla U(\bmtheta)$ required by the integrator, and the acceptance rate depends on the numerical integration error, the choice of the integration method is key to the efficiency of the HMC algorithm.

\section{Multi-stage integrators and adaptive approach}\label{sec:MultiStageIntegrators}
In this Section, we review multi-stage palindromic splitting integrators, which have demonstrated promising performance in HMC for both statistical and molecular simulation applications \cite{bcss2014, MAIApaper2017, AIApaper2016, bonilla2022interfacial, campos_sanz-serna2017,calvo2021hmc, bonillaetal2021, bonilla2021batteries}.

\subsection{k-stage palindromic splitting integrators}
 The family of palindromic $k$-stage splitting integrators with $k-1$ free parameters is defined as \cite{bou-rabee_sanz-serna_2018}
\begin{equation}\label{eq:kStageSchemeEven}
\Psi_h = \varphi^B_{b_1 h} \circ \varphi^A_{a_1 h} \circ \dots \circ \varphi^A_{a_{k'} h} \circ \varphi^B_{b_{k'+1} h} \circ \varphi^A_{a_{k'} h} \circ \dots \circ \varphi^A_{a_1 h} \circ \varphi^B_{b_1 h}, \quad b_i, a_j \in \mathbb{R}^+,
\end{equation}
if $k = 2 k'$, and
\begin{equation}\label{eq:kStageSchemeOdd}
\Psi_h = \varphi^B_{b_1 h} \circ \varphi^A_{a_1 h} \circ \dots \circ \varphi^B_{b_{k'} h} \circ \varphi^A_{a_{k'} h} \circ \varphi^B_{b_{k'} h} \circ \dots \varphi^A_{a_1 h} \circ \varphi^B_{b_1 h}, \quad b_i, a_j \in \mathbb{R}^+,
\end{equation}
if $k = 2k'-1$. The coefficients $b_i$, $a_j$ in \eqref{eq:kStageSchemeEven}-\eqref{eq:kStageSchemeOdd} have to satisfy the conditions $2 \sum_{i=1}^{k'} b_i + b_{k'+1} = 2 \sum_{j=1}^{k'} a_j = 1$, and $2 \sum_{i=1}^{k'} b_i = 2 \sum_{j=1}^{k'-1} a_j + a_{k'}= 1$, respectively. The integrators \eqref{eq:kStageSchemeEven} and \eqref{eq:kStageSchemeOdd} are symplectic as compositions of flows of Hamiltonian systems, and reversible, due to their palindromic structure.
The number of stages $k$ is the number of times the algorithm performs an evaluation of gradients $\nabla_{\theta} U(\bmtheta)$ per step size. Though $\varphi^B$ appears $k+1$ times in \eqref{eq:kStageSchemeEven} and \eqref{eq:kStageSchemeOdd}, the number of gradient evaluations performed is still $k$ since the (last) one in the leftmost $\varphi^B_{b_1 h}$ at the current step is reused in the rightmost $\varphi^B_{b_1 h}$ at the following step.
Multi-stage splitting integrators alternate position drifts and momentum kicks of different  lengths, which makes all of them, including the most common and popular 1-stage Verlet \eqref{eq:1sVVmap}, easy to implement.

As pointed out above, most of the computational effort in HMC is due to evaluations of gradients. Splitting integrators with different numbers of stages do not perform the same number of gradient evaluations per integration step and therefore using those integrators with a common value of $L$ and $h$ does not result in fair comparisons (in terms of computational cost). If $\hat{L}$ is a number of gradient evaluations/time steps suitable for the $1$-stage Verlet algorithm with step size $h$, $k$-stage integrators will here be used by taking
$L=\hat{L}/k$ steps of length $k h$. In this way all algorithms integrate the Hamiltonian dynamics over a time interval of the same length $\hat{L}h$ and use the same number of gradient evaluations.

\subsection{Examples of 2- and 3-stage integrators}\label{sec:ExamplesTwoThreeStageSchemes}
We plan to derive adaptive 2- and 3-stage integrators and we first review the examples in the literature of 2- and 3-stage integrators.

The one-parameter family of 2-stage integrators is described as (see \eqref{eq:kStageSchemeEven}):
$$
\Psi^{\text{2stage}}_h = \varphi^B_{b h} \circ \varphi^A_{a h} \circ \varphi^B_{b_1 h} \circ \varphi^A_{a h} \circ \varphi^B_{b h},
$$
with $a = 1/2$ and $b_1 = 1-2b$.
 Thus the integrators can be written as
\begin{equation}\label{eq:2stageIntegrators}
\Psi^{\text{2stage}}_h = \varphi^B_{b h} \circ \varphi^A_{\frac{h}{2}} \circ \varphi^B_{\left( 1 - 2b \right) h} \circ \varphi^A_{\frac{h}{2}} \circ \varphi^B_{b h},
\end{equation}
with $b \in (0, 0.5)$ if we wish $b>0$ and $b_1>0$.

Similarly, \eqref{eq:kStageSchemeOdd} with $k' = 2$, $2a + a_1 = 1$ and $2b + 2b_1 = 1$ yields the two-parameter family of $3$-stage integrators
\begin{equation}\label{eq:3stageIntegrators}
\Psi^{\text{3stage}}_h = \varphi^B_{b h} \circ \varphi^A_{a h} \circ \varphi^B_{\left( \frac{1}{2} - b \right) h} \circ \varphi^A_{\left( 1 - 2a \right) h} \circ \varphi^B_{\left( \frac{1}{2} - b \right) h} \circ \varphi^A_{a h} \circ \varphi^B_{b h},
\end{equation}
with $a, b \in (0, 0.5)$.

Several 2- and 3-stage integrators with suitably chosen parameters for achieving high performance in HMC have been proposed in the literature \cite{mclachlan1995, takaishi_deforcrand_2006, bcss2014, campos_sanz-serna2017}. Some of them are presented below and summarized  in Table~\ref{tab:IntegratorsTable}. In the cited literature, two alternative types of analysis have been carried out in order to choose the integration parameters $a$ and/or $b$ in the context of HMC. In \cite{mclachlan1995, takaishi_deforcrand_2006} or \cite{mclachlan_atela1992}, the integration coefficients are determined by minimizing the coefficients in the Taylor expansion of the \emph{Hamiltonian truncation error} \cite{mclachlan_atela1992}
\begin{equation}\label{eq:HamiltonianTruncationError}
\epsilon = H(\bmtheta, \bmp) - H(\Psi_h (\bmtheta, \bmp)).
\end{equation}
On the other hand, the paper \cite{bcss2014} does not look at the behavior of the Hamiltonian truncation error as $h\rightarrow 0$, as typically integrators are not operated with small values of $h$. Their analysis is rather based on  a (tight) bound
\begin{equation*}
\mathbb{E}[\Delta H] \le \rho(h, \bmz),
\end{equation*}
for the expected energy error with respect to $\pi(\bm{\theta}, \bm{p})$ \eqref{eq:InvariantJointDistribution} for given $h$, that may be rigorously proved for Gaussian targets (and has been experimentally shown to be useful for all targets). Here $\Delta H$ is an energy error as defined in \eqref{eq:EnergyError}, $\rho$ is a function associated with the integrator and $\bmz$ represents the coefficients that identify the integrator within a family.
For 2-stage palindromic splitting schemes \cite{bcss2014}
\begin{equation}\label{eq:rho2stage}
\rho_2 (h, b) = \frac{h^4 \left( 2b^2 \left( \frac{1}{2}-b \right) h^2 + 4b^2 - 6b + 1 \right)^2}{8 \left(2 - b h^2 \right) \left(2- \left( \frac{1}{2} - b \right) h^2 \right) \left(1 - b \left( \frac{1}{2} - b \right) h^2 \right)}.
\end{equation}
For 3-stage integrators the attention may be restricted to pairs $(b, a)$ that satisfy
\cite{campos_sanz-serna2017, radivojevic2018}
\begin{equation}\label{3stageHyperbola}
6 a b - 2 a - b + \frac{1}{2} = 0;
\end{equation}
when this condition is not fulfilled the integrator has poor stability properties \cite{campos_sanz-serna2017}.
Under this restriction (see \ref{app:rho3stage})
\begin{equation}\label{eq:rho3stage}
\rho_3 (h, b) =
\scalebox{0.95}{$ \frac{h^4 \left( -3 b^4 + 8 b^3 -19/4 b^2 + b + b^2 h^2 \left( b^3 - 5/4 b^2 + b/2 - 1/16 \right) - 1/16 \right)^2}{2 \left( 3 b - b h^2 \left( b - 1/4 \right) - 1 \right) \left(1 - 3 b - b h^2 \left(b - 1/2 \right)^2 \right) \left( -9 b^2 + 6 b - h^2 \left( b^3 - 5/4 b^2 + b/2 - 1/16 \right) - 1 \right)}$}.
\end{equation}

The following schemes have been considered in the literature.
\begin{itemize}
\item \textbf{2-stage Velocity Verlet (VV2).}
This is the integrator with the longest stability interval $(0, 4)$ for an integration step size $h$ among 2-stage splitting schemes and corresponds to $b = 1/4$ in \eqref{eq:2stageIntegrators}.
To perform one step of length $h$ with this algorithm, one just performs two steps of length $h/2$ of standard Velocity Verlet.
It means that performance comparison of alternative 2-stage splitting integrators with standard Velocity Verlet can be achieved through comparison with VV2 if the step length and number of steps per integration leg are adjusted accordingly.

\item \textbf{2-stage BCSS (BCSS2).}
This scheme was derived in \cite{bcss2014} to minimize the maximum of $\rho_2 (h, b)$ in \eqref{eq:rho2stage} as $h$ ranges over the interval $0<h<2$ (VV2 is often operated with $h$ close to $2$), i.e.\
\begin{equation*}
b = \argmin_{b \in \left(0, 0.5 \right) } \max_{0 < h < 2} \rho_2 (h, b) = 0.211781.
\end{equation*}
It achieves its best performance when $h$ is near the center of the longest stability interval, i.e. $h \approx 2$ \cite{bcss2014, MAIApaper2017, AIApaper2016, radivojevic2018, mazur1997, mazur1998}.

\item \textbf{2-stage Minimum Error (ME2).}
The coefficient of this integrator ($b = 0.193183$) was obtained by McLachlan in \cite{mclachlan1995} through the minimization of the Hamiltonian truncation error \eqref{eq:HamiltonianTruncationError}. For quadratic problems, see also \cite{radivojevic2018}.

\item \textbf{3-stage Velocity Verlet (VV3).}
Similarly to VV2, the 3-stage Velocity Verlet is a 3-stage integrator with the longest stability interval $(0, 6)$ among 3-stage splitting integrators.  One step of this algorithm of length $h$ is just the concatenation of three steps of length $h/3$ of the standard Velocity Verlet integrator. As we did for VV2, we emphasize that when comparing below alternative integrators with VV3, one is really comparing them with the standard VV algorithms.

\item \textbf{3-stage BCSS (BCSS3).}
The parameter values are found by imposing the relation \eqref{3stageHyperbola} and
\begin{equation*}
b = \argmin_{b \in \left(0, 0.5 \right) } \max_{0 < h < 3} \rho_3 (h, b),
\end{equation*}
with $\rho_3$ in \eqref{eq:rho3stage}.

\item \textbf{3-stage Minimum Error (ME3).}
ME3 was derived in \cite{predescu2012}  by requiring \eqref{3stageHyperbola} and a Hamiltonian truncation error of size $\mathcal{O} (h^6)$.
\end{itemize}
\begin{table}[!ht]
\centering
\resizebox{\textwidth}{!}{
\begin{tabular}{c c c c c}
Integrator & N. of stages & Coefficients & Stability interval & References \\
\hline
Velocity Verlet & $1$ & - & $(0, 2)$ & \cite{verlet1967, swope1982} \\
2-stage Velocity Verlet & $2$ & $b = 1/4$ & $(0, 4)$ & \cite{bcss2014} \\
2-stage BCSS & $2$ & $b = 0.211781$ & $(0, 2.634)$ & \cite{bcss2014} \\
2-stage Minimum Error & $2$ & $b = 0.193183$ & $(0, 2.533)$ & \cite{mclachlan1995, radivojevic2018} \\
3-stage Velocity Verlet & $3$ & $b = 1/6$, $a = 1/3$ & $(0, 6)$ & \cite{bcss2014, campos_sanz-serna2017} \\
\multirow{2}{*}{3-stage BCSS} & \multirow{2}{*}{$3$} & $b = 0.118880$ & \multirow{2}{*}{$(0, 4.662)$} & \multirow{2}{*}{\cite{bcss2014, campos_sanz-serna2017}} \\
& & $a = 0.296195$ & & \\
\multirow{2}{*}{3-stage Minimum Error} & \multirow{2}{*}{$3$} & $b = 0.108991$ & \multirow{2}{*}{$(0, 4.584)$} & \multirow{2}{*}{\cite{predescu2012, campos_sanz-serna2017}} \\
& & $a = 0.290486$ & & \\
\hline
\end{tabular}
}
\caption{\label{tab:IntegratorsTable} Multi-stage splitting integrators presented in Section \ref{sec:ExamplesTwoThreeStageSchemes}.}
\end{table}

The performance of the different integrators within HMC very much depends on the simulation parameters, in particular on the choice of step size.
 Minimum Error schemes achieve their best performance for small step size, since they are obtained by studying the limit of vanishing step size. However, they have shorter \emph{stability limits}, i.e. lengths of the stability intervals, and may perform badly for bigger integration step sizes. Velocity Verlet schemes preserve stability for values of the step size larger than those that may be used in other integrators, but may not be competitive in situations where the step size is not chosen on grounds of stability (for instance in problems of large dimensionality where accuracy demands that the step size be small to ensure non-negligible acceptance rates).
 BCSS integrators were designed for optimizing performance for values of the step size not close to 0 and not close to the maximum stability allowed for Verlet.

\subsection{Adaptive Integration Approach (AIA)}
\begin{sloppypar}
Adaptive 2-stage integration schemes were proposed by Fernández-Pendás et al. in \cite{AIApaper2016} for molecular simulation applications. Their extensions, called MAIA and e-MAIA, for Modified HMC (MHMC) methods, such as Generalized Shadow HMC (GSHMC) methods \cite{GSHMCpaper2008, escribano2015MTS_GSHMC, akhmatskaya_reich2011, akhmatskaya2011mesoGSHMC}, were introduced by Akhmatskaya et al. in \cite{MAIApaper2017}.
\end{sloppypar}

Given a simulation problem, in AIA, the user chooses, according to their computational budget, the value of $h$ to be used (i.e. $h$ is chosen to be smaller if more time and resources are available for the simulation). After that, the AIA algorithm itself finds the most appropriate integration scheme within the family of $2$-stage integrators \eqref{eq:2stageIntegrators}. If the time-step is very small for the problem at hand, AIA will automatically pick up a parameter value close to Minimum Error; if the time-step is very large, AIA will automatically choose an integrator close to the $2$-stage Velocity Verlet. For intermediate values of $h$, AIA will choose an intermediate parameter value (near the BCSS integrator). We emphasize that in AIA, the parameter value used changes with $h$ and with the problem being tackled.
Given a simulation problem, the AIA offers, for any integration step size chosen within an appropriate stability  interval, an intelligent choice of the most appropriate integration scheme (in terms of the best conservation of energy for harmonic forces)  within a family of 2-stage integrators.
The original AIA algorithm is summarized in Algorithm~\ref{alg:AIA}.

Our objective in this paper is to employ the ideas behind the 2-stage AIA approach for deriving multi-stage adaptive integration schemes specifically addressed to Bayesian inference applications. Taking into account the recent indications of the superiority of $3$-stage integrators over $2$-stage schemes in statistical applications \cite{radivojevic2018}, we plan to develop not only  $2$-stage adaptive approaches as in AIA but also $3$-stage adaptive algorithms.
Extending AIA to computational statistics is not straightforward. The potential challenges are discussed in the next Section.
\begin{algorithm}
\captionsetup{labelsep=AIAalgorithm}
\hrulefill
\begin{algorithmic}[1]
\Input highest angular frequency $\tilde{\omega}$ in the problem, dimensional time-step $\overline{\Delta t}$, safety factor $S_{{f}_{\text{AIA}}} = \sqrt{2}$ \cite{AIApaper2016}
\State Calculate dimensionless time-step: $\overline{h} \gets S_{{f}_{\text{AIA}}} \tilde{\omega} \overline{\Delta t}$
\If {$\overline{h} \ge 4$}
	\State \textbf{abort} - there does not exist an integration coefficient $b$ for which a $2$-stage integrator $\Psi^{\text{2stage}}_h$ in \eqref{eq:2stageIntegrators} is stable
\Else
\State Find optimal integrator coefficient:
\Statex $\displaystyle b_{\text{opt}} \gets \argmin_{0 < b < 0.5} \max_{0 < h < \overline{h}} \rho_2 (h, b)$
\EndIf
\Output An integration coefficient $b_{\text{opt}}$ which determines an optimal 2-stage integrator $\Psi^{\text{2stage}}_h$ in \eqref{eq:2stageIntegrators} to be used in an HMC simulation of the given physical system with integration time-step $\overline{\Delta t}$
\end{algorithmic}
\hrulefill
\caption{Adaptive Integration Approach (AIA). Given a physical system and a time-step $\overline{\Delta t}$, AIA offers the most appropriate choice of an integration parameter $b$ for a 2-stage splitting integrator \eqref{eq:2stageIntegrators}.}\label{alg:AIA}
\end{algorithm}

\section{s-AIA}\label{sec:sAIA}
\subsection{Extension of AIA to computational statistics}\label{sec:extensionAIAtoCompStat}
AIA makes use of specific properties and assumptions that hold for molecular simulation problems, e.g.\ the strongest forces in the target distribution are approximately harmonic (Gaussian) with known angular frequencies, there are well determined \emph{safety factors} which scales the longest integration stability interval to avoid nonlinear resonances, and the step size does not vary from one integration leg to the next. Unfortunately, those conditions are not usually met in Bayesian inference applications and therefore, when formulating s-AIA, the statistics version of AIA, the following issues have to be dealt with.
\begin{itemize}	
\item \textbf{Harmonic forces.}
In contrast to molecular systems, they are not typically dominating in the Bayesian scenario.
\item \textbf{Computation of frequencies.} Even if the integrator could be chosen by examining only harmonic forces, the corresponding
angular frequencies would not be known  a priori in a Bayesian simulation.
\item \textbf{Resonance conditions.}
Restrictions on the integration step size imposed by nonlinear stability are not known in the Bayesian case.
\item \textbf{Choice of a step size.}
In statistics, the step size is usually randomized at the beginning of each integration leg and this would involve having to adjust at each step of the Markov chain the parameter values within the chosen family of integrators  (see Step 5 in Algorithm~\ref{alg:AIA}).
\end{itemize}

We address these issues separately.

\subsubsection*{Pre-tabulation of the map $\overline{h} \to b_{\text{opt}}$}
For each family of methods  ($2$- or $3$-stage), we tabulate \textit{once and for all} the optimal integration coefficients $b_{\text{opt}}^k$, $k = 2, 3$, at small increments of $\overline{h}$ \cite{sAIA_tables}. In this way, the extra computational effort due to Step 5 in Algorithm~\ref{alg:AIA} can be avoided.

We produced  tables for  $k$-stage s-AIA, $k = 2, 3$, using  grids $\{ \overline{h}_i \}_k$, $i = 1, ... , N_{\text{grid}}$ of the dimensionless stability interval $(0, 2 k)$ ($N_{\text{grid}}$ controls the accuracy of the estimated $b_{\text{opt}}^k$ for a given $\overline{h}$). Similarly to Algorithm~\ref{alg:AIA}, $\{ b_{{\text{opt}}_i}^k \}$, $i = 1, ... , N_{\text{grid}}$, $k = 2, 3$, are found as
\begin{align}\label{eq:boptsAIA}
b_{{\text{opt}}_i}^k = \argmin_{b \in \left( b_{\text{ME}k}, \, b_{\text{VV}k} \right) }  \max_{0 < h < \overline{h_i}} \rho_k (h, b), \\
\overline{h_i} \in \{ \overline{h}_i \}_k, \quad i = 1, ... , N_{\text{grid}}, \quad k = 2, 3, \nonumber
\end{align}
where $b_{\text{ME} k}$ (the optimal parameter for the $k$-stage integrator as $h \to 0$) and $b_{\text{VV} k}$ (the longest stability limit for the $k$-stage family) are the boundaries for $b$, and $\rho_2 (h, b)$, $\rho_3 (h,b)$ are given by \eqref{eq:rho2stage} and \eqref{eq:rho3stage} respectively. For 3-stage s-AIA, the second parameter $a$ in \eqref{eq:3stageIntegrators} is calculated according to \eqref{3stageHyperbola}.

\begin{figure}[ht!]
\centering
\includegraphics[width = \textwidth]{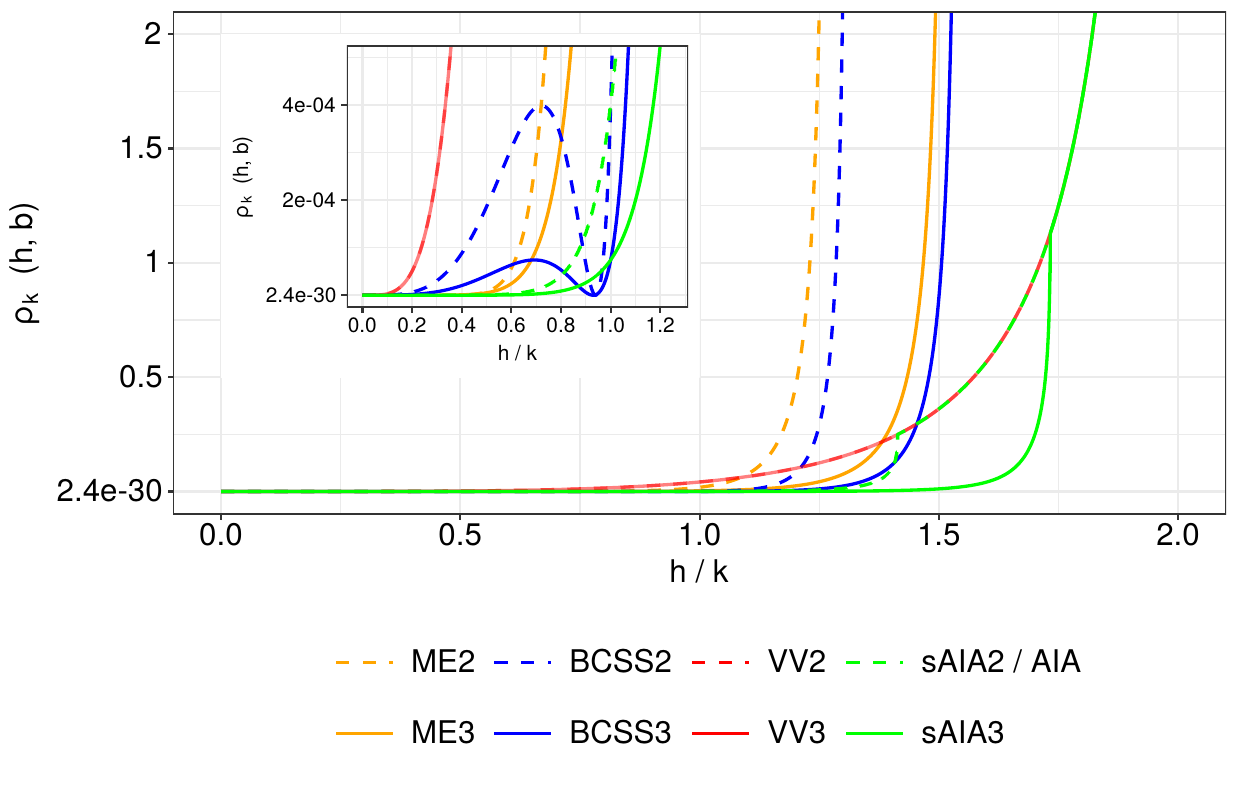}
 \caption{Comparison of the upper bounds  $\rho_k (h, b)$, $k = 2$ \eqref{eq:rho2stage}$,3$ \eqref{eq:rho3stage} of the energy error, for fixed-parameter multi-stage splitting integrators --- VV2, VV3, BCSS2, BCSS3, ME2, ME3 (Table \ref{tab:IntegratorsTable}) --- and the adaptive integrators AIA and s-AIA$k$. The interval for the step size $h$ is normalized with respect to the number of stages $k$ of the integrator in order to lead to fair comparisons. The zoomed plot in the upper left corner shows the situation for $h/k \in (0, 1.2)$.}
 \label{fig:rho_comparison}
\end{figure}

\begin{figure}[ht!]
\centering
\includegraphics[width = \textwidth]{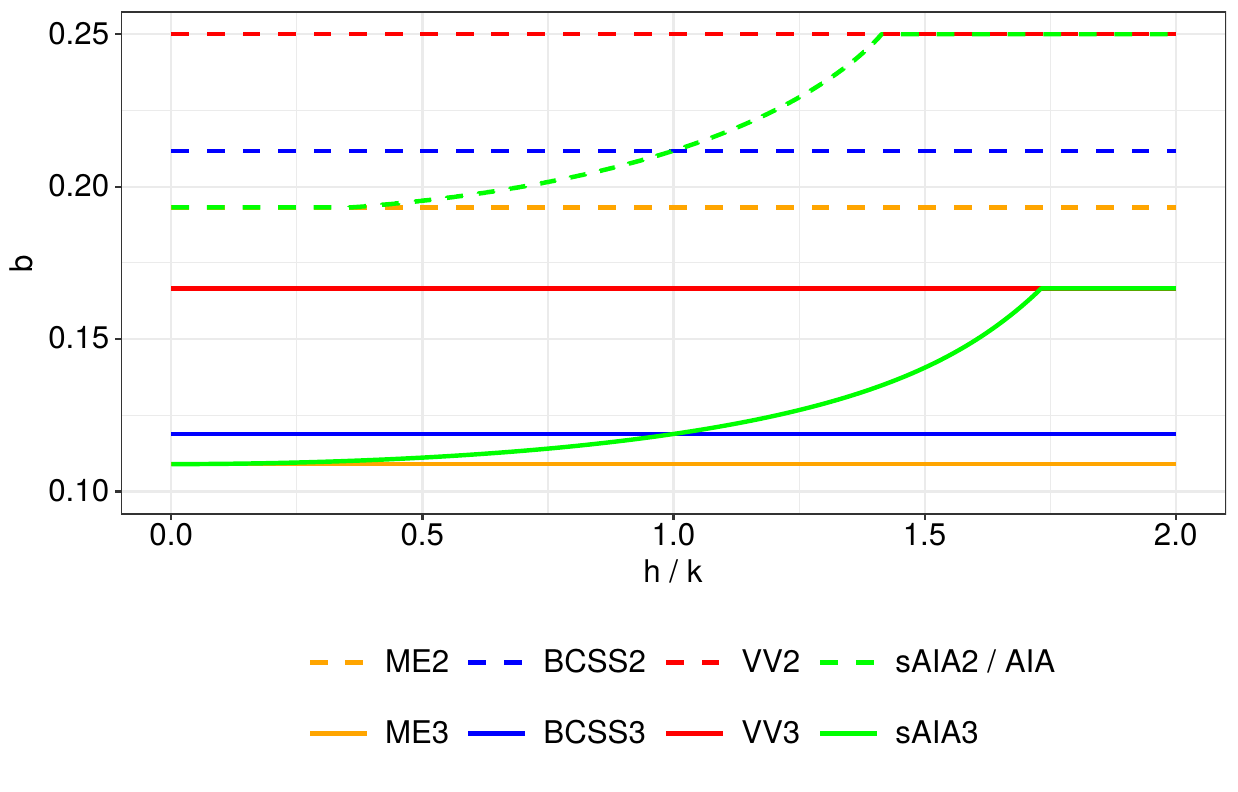}
 \caption{Comparison of the integration coefficient $b$ for fixed-parameter multi-stage splitting integrators ---VV2, VV3, BCSS2, BCSS3, ME2, ME3 (Table \ref{tab:IntegratorsTable})--- and the adaptive integrators AIA and s-AIA$k$, $b^2_{\text{opt}}$ and $b^3_{\text{opt}}$ \eqref{eq:boptsAIA}. The interval for the step size $h$ is normalized with respect to the number of stages $k$ of the integrator to lead to fair comparisons. }
 \label{fig:b_comparison}
\end{figure}

Similarly to what happens in AIA, in  s-AIA, one expects $b_{\text{opt}}^k$ to be close to the ME$k$ integrator coefficients for smaller values of $h$; to be close to $b_{\text{BCSS}k}$ near $\overline{h} = k$, and to increase up to $b_{\text{VV}k}$ as $\overline{h}$ approaches $2k$.
Figure~\ref{fig:rho_comparison} shows the $\rho_2 (h, b)$ and $\rho_3 (h, b)$ functions for the range of adaptive and fixed-parameter multi-stage integrators discussed in this work, whereas Figure~\ref{fig:b_comparison} depicts $b^2_{\text{opt}}$ and $b^3_{\text{opt}}$ as functions of dimensionless step size.

\subsubsection*{Computation of frequencies}
The frequencies $\omega_j$, $j = 1, ... , D$, of the system are calculated during the burn-in stage (a
mandatory initial stage of an HMC simulation to reach its stationary regime) as
\begin{equation}\label{eq:Frequencies}
\omega_j = \sqrt{\lambda_j}, \qquad j = 1, ... , D,
\end{equation}
where $\lambda_j$ are the eigenvalues of the Hessian matrix of the potential function
\begin{equation*}
H_{i, j} = \frac{\partial^2 U(\bmtheta)}{\partial \theta_i \partial \theta_j}, \qquad i, j = 1, ... , D.
\end{equation*}
Since the Hessian matrix evolves during an HMC simulation, the resulting frequencies are calculated as averages of \eqref{eq:Frequencies} over the burn-in stage.

\subsubsection*{Calculation of fitting factors}\label{sec:sAIAFittingFactor}
Explicit integrators, such as the ones discussed in this study, may become unstable, and thus suffer from  serious step size limitations when applied to nonlinear Hamiltonian systems \cite{schlick_etal1998}. To quantify the step size limitations imposed by nonlinear stability in the Verlet integrator, Schlick et al. \cite{schlick_etal1998} introduced
stability limits on $\omega \Delta t$ for up to the 6th order resonances. This seemed to cover the worst scenarios in molecular simulations.
On the other hand, to reproduce angular frequencies in the presence of nonlinear resonances, the authors of \cite{AIApaper2016} proposed to multiply them by the so called safety factor (SF). This also can be interpreted as a reduction of the stability limit by SF times. The safety factors are closely related to the stability limits on $\omega \Delta t$ provided in Table 1 in \cite{schlick_etal1998} for the range of resonance orders. In particular, $\text{SF} = \frac{2}{\omega \Delta t}$. For readers' convenience, we list the values of SF for the Verlet integrator that correspond to the resonance orders ranging from $2$ to $6$ in Table \ref{tab:SafetyFactors}.
We have already mentioned that AIA \cite{AIApaper2016} makes use of a safety factor $\sqrt{2}$ (cf. Algorithm~\ref{alg:AIA}), which avoids resonances up to  $4$-th order, while the MAIA algorithm for Modified HMC \cite{MAIApaper2017} utilizes $\sqrt{3}$, that covers resonances up to  $5$-th order.
In Bayesian inference applications, the number of multiple time scales and the level of non-linearity are in general hardly predictable, and should be treated for each problem separately. For our purposes, instead of a safety factor, we introduce what we call a fitting factor $S_f$, which not only plays the role of the safety factor  but also results from fitting the proposed multivariate Gaussian model to the data generated during the burn-in stage. As in the case of a safety factor in \cite{AIApaper2016}, we use a fitting factor for nondimensionalization of the step size. Thus, for a chosen step size $\overline{\Delta t}$, its nondimensional counterpart is found as
\begin{equation}\label{eq:Nondimensionalization}
\overline{h} = S_f \, \tilde{\omega} \, \overline{\Delta t}.
\end{equation}
\begin{table}[!ht]
\centering
\begin{tabular}{c c}
Resonance order & Safety factor \\
\hline
$2$ & $1$ \\
$3$ & $\frac{2}{3}\sqrt{3} \approx 1.15$ \\
$4$ & $\sqrt{2} \approx 1.41$ \\
$5$ & $\frac{\sqrt{5}\sqrt{10 + 2 \sqrt{5}}}{5} \approx 1.70$ \\
$6$ & $2$ \\
\hline
\end{tabular}
\caption{\label{tab:SafetyFactors} Safety factor values for avoiding resonances of up to the $6^{\text{th}}$ order.}
\end{table}
\\
Here, $S_f$ is the fitting factor determined below and $\tilde{\omega}$ is the highest frequency of the system, obtained from the burn-in simulation. Our objective now is to express $S_f$ in terms of the known properties of the simulated system.
We choose to run a burn-in simulation using a Velocity Verlet algorithm and setting $L = 1$ and $\Delta t = \Delta t_{\text{VV}}$, where $\Delta t_{\text{VV}}$ is an integration step size properly adjusted to reach a user-chosen target acceptance rate $\alpha_{\text{target}}$. In \ref{app:ARtarget}, we provide our recommendation for selecting an $\alpha_{\text{target}}$ which yields the highest level of accuracy for fitting factor and frequencies estimations within the reasonable computational time. The choice of Verlet with $L = 1$ helps to obtain a simple closed-form expression for the expected energy error $\mathbb{E} [\Delta H]$ (see \ref{app:UnivEnergyErrorExpectation1sVV} for details):
%
\begin{equation}\label{eq:UnivariateEnergyErrorExpectation}
\mathbb{E}^1_{\text{VV}} [\Delta H] = \frac{h_{\text{VV}}^6}{32},
\end{equation}
with $h_{\text{VV}}$ being a dimensionless counterpart of $\Delta t_{\text{VV}}$, i.e. from \eqref{eq:Nondimensionalization}
\begin{equation*}
h_{\text{VV}} = S_f \, \omega \, \Delta t_{\text{VV}} .
\end{equation*}
For a $D$-dimensional multivariate Gaussian target, one can consider $D$ dimensionless counterparts
\begin{equation}\label{eq:MultivariatehVV}
h_{{\text{VV}}_j} = S_f \, \omega_j \, \Delta t_{\text{VV}}, \qquad j = 1, ... , D ,
\end{equation}
and find the expected energy error for a multivariate Gaussian model with the help of \eqref{eq:UnivariateEnergyErrorExpectation} as
\begin{equation}\label{eq:MultivariateEnergyErrorExpectation}
\mathbb{E}^D_{\text{VV}} [\Delta H] = \sum_{j=1}^D \frac{h^6_{{\text{VV}}_j}}{32}.
\end{equation}
Combining \eqref{eq:MultivariateEnergyErrorExpectation} and \eqref{eq:MultivariatehVV}, we find the fitting factor
\begin{equation}\label{eq:FittingFactorFrequencies}
S_f = \frac{1}{\Delta t_{\text{VV}}} \sqrt[6]{\frac{32 \mathbb{E}^D_{\text{VV}} [\Delta H]}{\sum_{j=1}^D \omega_j^6}}.
\end{equation}

Alternatively, the calculation of the frequencies may be avoided (and computational resources  saved), if the multivariate Gaussian model is replaced with a univariate Gaussian model (as in \cite{AIApaper2016}), which leads to
\begin{equation}\label{eq:FittingFactor}
S_f = \frac{1}{\tilde{\omega} \Delta t_{\text{VV}}} \sqrt[6]{\frac{32 \mathbb{E}^D_{\text{VV}} [\Delta H]}{D}}.
\end{equation}
Notice that, though $\tilde{\omega}$ appears in \eqref{eq:FittingFactor}, one can compute
\begin{equation}\label{eq:Somegatilde}
S_f \, \tilde{\omega} = \frac{1}{\Delta t_{\text{VV}}} \sqrt[6]{\frac{32 \mathbb{E}^D_{\text{VV}} [\Delta H]}{D}},
\end{equation}
without needing frequencies and use it in \eqref{eq:Nondimensionalization}.

From now on, in order to distinguish between the two approaches, we will denote the one in \eqref{eq:FittingFactorFrequencies} --- which requires frequency calculation --- by $S_{\omega}$ and the second one in \eqref{eq:FittingFactor} --- which does not --- by $S$, i.e.
\begin{equation*}
S_{\omega} = \frac{1}{\Delta t_{\text{VV}}} \sqrt[6]{\frac{32 \mathbb{E}^D_{\text{VV}} [\Delta H]}{\sum_{j=1}^D \omega_j^6}}, \qquad\qquad S = \frac{1}{\tilde{\omega} \Delta t_{\text{VV}}} \sqrt[6]{\frac{32 \mathbb{E}^D_{\text{VV}} [\Delta H]}{D}}.
\end{equation*}
As pointed out above, safety factors are meant to impose limitations on a system-specific stability interval (cf. \eqref{eq:Nondimensionalization}). Thus, they should not be less than $1$ and, as a consequence, we actually use
\begin{equation}\label{eq:FittingFactorsBothAsMax}
\scalebox{1}{$ S_{\omega} = \max \left(1, \frac{1}{\Delta t_{\text{VV}}} \sqrt[6]{\frac{32 \mathbb{E}^D_{\text{VV}} [\Delta H]}{\sum_{j=1}^D \omega_j^6}} \right), \, \, S = \max \left(1, \frac{1}{\tilde{\omega} \Delta t_{\text{VV}}} \sqrt[6]{\frac{32 \mathbb{E}^D_{\text{VV}} [\Delta H]}{D}} \right). $}
\end{equation}
We remark that, for $S_f$ in \eqref{eq:FittingFactor} smaller than 1, $S$ in \eqref{eq:FittingFactorsBothAsMax} is equal to $1$, then $\tilde{\omega}$ is required for a nondimensionalization as in \eqref{eq:Nondimensionalization}. However, following \cite{lecun1992}, $\tilde{\omega}$ can be computed avoiding the calculations of Hessians, i.e. without introducing a computational overhead.

The only unknown quantity in \eqref{eq:FittingFactorsBothAsMax} is $\mathbb{E}^D_{\text{VV}} [\Delta H]$, which can be found by making use of the data collected during the burn-in stage. In fact,
following the high-dimensional asymptotic formula for expected acceptance rate $\mathbb{E} [\alpha]$ \cite{beskos_optimal_tuning} proven for Gaussian distributions in a general scenario \cite{calvo2021hmc}, i.e.
\begin{equation*}
\mathbb{E} [\alpha] = 1 - \frac{1}{2\sqrt{\pi}} \sqrt{\mathbb{E}^D [\Delta H]}, \qquad \mathbb{E}^D [\Delta H] \to 0, \, D \to \infty,
\end{equation*}
we get an approximation for $\mathbb{E}^D [\Delta H]$
\begin{equation}\label{eq:EdHBeskosApproach}
\mathbb{E}^D [\Delta H] \, \approx \, 4 \pi \left( 1 - \mathbb{E} [\alpha] \right)^2 .
\end{equation}
An estimation of $\mathbb{E} [\alpha]$ in a simulation is given by the acceptance rate AR, i.e. the ratio between the accepted $N_{\text{acc}}$ and the total $N$ number of proposals
\begin{equation}\label{eq:ARcalculation}
\text{AR} = \frac{N_{\text{acc}}}{N}.
\end{equation}
Combining \eqref{eq:EdHBeskosApproach} with $\mathbb{E} [\alpha] = \text{AR}$ calculated during the burn-in stage, we compute $\mathbb{E}^D_{\text{VV}} [\Delta H]$ as
\begin{equation*}
\mathbb{E}^D_{\text{VV}} [\Delta H] = 4 \pi \left( 1 - \text{AR} \right)^2,
\end{equation*}
which gives an explicit expression for the fitting factors in \eqref{eq:FittingFactorsBothAsMax}
\begin{equation}\label{eq:FittingFactorsExplicit}
\scalebox{1}{$ S_{\omega} = \max \left(1, \frac{2}{\Delta t_{\text{VV}}} \sqrt[6]{\frac{2 \pi (1 - \text{AR})^2}{\sum_{j=1}^D \omega_j^6}} \right), \, \, S = \max \left(1, \frac{2}{\tilde{\omega} \Delta t_{\text{VV}}} \sqrt[6]{\frac{2 \pi (1 - \text{AR})^2}{D}} \right). $}
\end{equation}
Once the fitting factor is computed using \eqref{eq:FittingFactorsExplicit}, a dimensionless counterpart of a given step size $\overline{\Delta t}$ can be calculated either as
\begin{equation}\label{eq:NondimensionalizationSfreqs}
\overline{h}_{\omega} =
\begin{cases}
\frac{2 \tilde{\omega} \overline{\Delta t}}{\Delta t_{\text{VV}}} \sqrt[6]{\frac{2 \pi (1 - \text{AR})^2}{\sum_{j=1}^D \omega_j^6}}, \quad &\text{if } S_{\omega} > 1, \\
\tilde{\omega} \overline{\Delta t}, \quad &\text{otherwise},
\end{cases}
\end{equation}
or
\begin{equation}\label{eq:NondimensionalizationSnofreqs}
\overline{h} = 
\begin{cases}
\frac{2 \overline{\Delta t}}{\Delta t_{\text{VV}}} \sqrt[6]{\frac{2 \pi (1 - \text{AR})^2}{D}}, \quad &\text{if } S > 1, \\
\tilde{\omega} \overline{\Delta t}, \quad &\text{otherwise}.
\end{cases}
\end{equation}
We remark that for systems with disperse distributions of frequencies, i.e. when the standard deviation of frequencies, $\sigma$, is big, it might be useful to apply a nondimensionalization of $\overline{\Delta t}$ smoother than the proposed in \eqref{eq:NondimensionalizationSfreqs}.
In fact, a nondimensionalization method like in \eqref{eq:Nondimensionalization} cannot be able to properly catch the scattered frequencies of such systems. Therefore, for $\sigma > 1$, we propose to use a nondimensionalization
\begin{equation*}
\overline{h} = S_{\omega} \, \left(\tilde{\omega}-\sigma\right) \, \overline{\Delta t},
\end{equation*}
which brings to
\begin{equation}\label{eq:NondimensionalizationSfreqsStdev}
\overline{h}_{\omega} =
\begin{cases}
\frac{2\left( \tilde{\omega} - \sigma \right) \, \overline{\Delta t}}{\Delta t_{\text{VV}}} \sqrt[6]{\frac{2 \pi (1 - \text{AR})^2}{\sum_{j=1}^D \omega_j^6}}, \quad &\text{if } S_{\omega} > 1, \\
\left( \tilde{\omega} - \sigma \right) \, \overline{\Delta t} \quad &\text{otherwise}.
\end{cases}
\end{equation}
On the other hand, if $\sigma < 1$, \eqref{eq:NondimensionalizationSfreqs} is a better choice. We remark that the choice of a treshold $\sigma = 1$ for using a smoother normalization method is heuristic and validated by the good results obtained in the numerical experiments (Sec. \ref{sec:NumericalResults}), as well as by the fact that the small $\sigma$ implies the negligible difference between \eqref{eq:NondimensionalizationSfreqs} and \eqref{eq:NondimensionalizationSfreqsStdev}. The second statement follows from the inspection of the ratio of $\overline{h}_{\omega}$ in \eqref{eq:NondimensionalizationSfreqsStdev} to $\overline{h}_{\omega}$ in \eqref{eq:NondimensionalizationSfreqs}. In Section \ref{sec:sAIAalgorithm} we will analyze different choices of scaling and provide practical recommendations. With \eqref{eq:NondimensionalizationSfreqs}-\eqref{eq:NondimensionalizationSfreqsStdev} one has everything in place for finding the optimal integrator parameter $b_{\text{opt}}^k$ \eqref{eq:boptsAIA}.

To conclude this section, it is worth mentioning yet another useful output of the analysis. Let us recall that the dimensionless maximum stability limit of $k$-stage integrators is equal to $2 k$, $k = 1, 2, 3, ...$ \cite{bou-rabee_sanz-serna_2018}. Then, the stability  interval can be expressed in terms of the chosen fitting factor $S_f$ ($S$ or $S_{\omega}$ in \eqref{eq:FittingFactorsBothAsMax}) as $\big( 0, {2 k}/({S_f \tilde{\omega}} )\big)$, $k = 1, 2, 3, ...$, or
\begin{equation}\label{ineq:StabilityLimitCalculation}
0 < \Delta t < \text{SL} = \frac{2 k}{S_f \, \tilde{\omega}}, \qquad k = 1, 2, 3, ... \, .
\end{equation}
Here SL is the stability limit. We remark that, with the nondimensionalization \eqref{eq:NondimensionalizationSfreqsStdev}, the estimation of the stability  interval differs from \eqref{ineq:StabilityLimitCalculation} and reads as
\begin{equation}\label{ineq:StabilityLimitCalculationStdev}
0 < \Delta t < \text{SL} = \frac{2 k}{S_{\omega} \,  \left( \tilde{\omega} - \sigma \right)}, \qquad k = 1, 2, 3, ... \, .
\end{equation}

In summary, we have proposed an approach for the prediction of a stability interval and an optimal multi-stage integrator for a given system. The step size can be freely chosen within the estimated stability interval.

\subsection{s-AIA algorithm}\label{sec:sAIAalgorithm}
Since the nondimensionalization method forms a key part of the s-AIA algorithm, it is important to give some insight into the options offered by \eqref{eq:NondimensionalizationSfreqs}-\eqref{eq:NondimensionalizationSfreqsStdev}. Obviously, the method \eqref{eq:NondimensionalizationSnofreqs} is cheaper in terms of computational effort as it does not require the calculation  of frequencies. In addition, \eqref{eq:NondimensionalizationSnofreqs} is not affected by potential inaccuracies of the computed frequencies due, e.g., to insufficient sampling during the burn-in stage. On the other hand, taking into account the different frequencies (hence, the different time scales) of the system provides a more accurate estimation of the system-specific stability  interval. Moreover, in the case of dominating anharmonic forces, the analysis based on the univariate harmonic oscillator model may lead to poor estimation of the fitting factor $S$ and, as a result, of the dimensionless step size in \eqref{eq:NondimensionalizationSnofreqs}. Therefore, we expect $S_{\omega}$ in \eqref{eq:FittingFactorsExplicit} to provide a better approximation of the stability interval, and thus to lead to a better behavior of s-AIA.
However, with the upper bound of the safety factor for the 1-stage Velocity Verlet suggested in \cite{schlick_etal1998}, it is possible to identify those computational models for which the less computationally demanding fitting factor $S$ ensures a reliable stability limit estimation. In particular, $S > 2$ implies an anharmonic behavior of the underlying dynamics of the simulated model, and thus the need for a more accurate $S_{\omega}$, together with \eqref{eq:NondimensionalizationSfreqs} or \eqref{eq:NondimensionalizationSfreqsStdev} (depending on the distribution of $\omega_j$), for a proper estimation of the stability limit. On the contrary, if $S \le 2$, one expects $S$ and \eqref{eq:NondimensionalizationSnofreqs} to be able to provide a reliable approximation of the stability limit.
Though, in contrast to \eqref{eq:NondimensionalizationSnofreqs}, the calculation of $S$ in \eqref{eq:FittingFactorsExplicit} requires the knowledge of the highest frequency $\tilde{\omega}$, it is still less computationally demanding than the $S_{\omega}$ approach since $\tilde{\omega}$ can be computed avoiding calculations of Hessians \cite{lecun1992}, which is the bulk of computational cost for the frequencies calculations. We remark that the option to avoid calculating frequencies and use \eqref{eq:NondimensionalizationSnofreqs} straightaway is present in the s-AIA algorithm.

\begin{figure}[ht!]
	\centering
	\includegraphics[width = \textwidth]{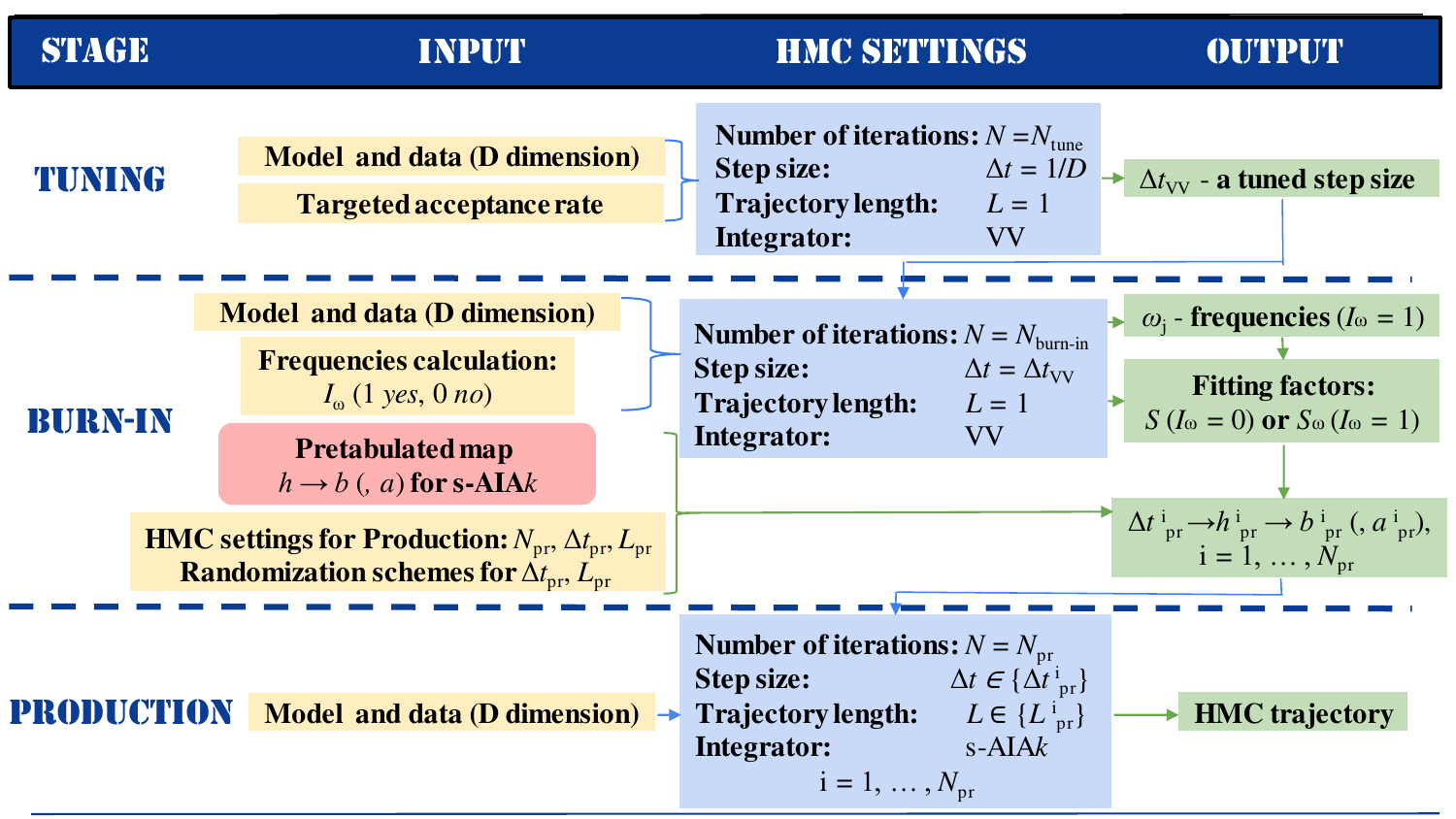}
	\caption{Summary of the s-AIA$k$ algorithm. The proposed approach consists of three stages: (i) tuning stage for adjusting the step size $\Delta t_{\text{VV}}$ to get $\text{AR} \approx \alpha_{\text{target}}$ (\ref{app:ARtarget}); (ii) burn-in stage; the optimal multi-stage integrator and the HMC simulation parameters are found by combining the simulation data and the analysis provided;
		(iii) production stage to generate the HMC samples.}
	\label{fig:sAIA_chart}
\end{figure}

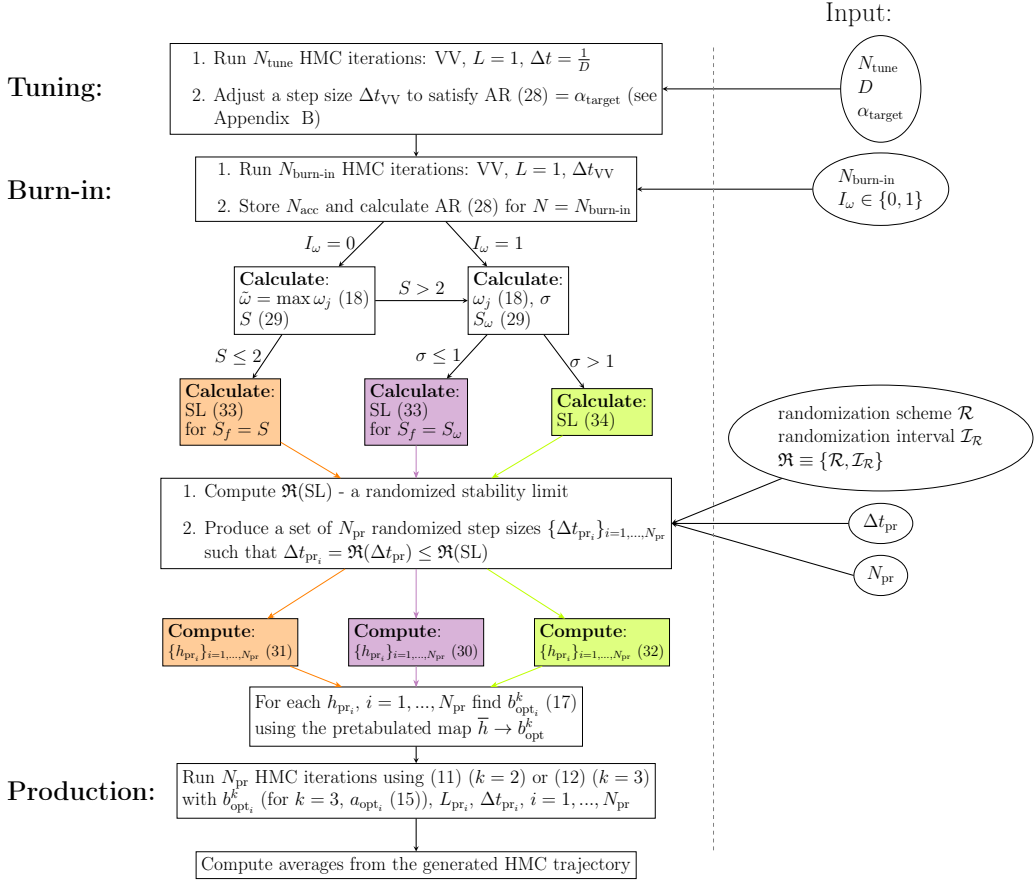
\begin{figure}[ht!]
\centering
{\resizebox{\textwidth}{!}{\begin{tikzpicture}[node distance=2cm]
\node (tuning_box) [draw, align=left, fill=white!30] {\large \begin{varwidth}{\linewidth}\begin{itemize}
\item[1.] Run $N_{\text{tune}}$ HMC iterations: VV, $L = 1$, $\Delta t = \frac{1}{D}$
\item[2.] Adjust a step size $\Delta t_{\text{VV}}$ to satisfy AR \eqref{eq:ARcalculation} $= \alpha_{\text{target}}$ (see \ref{app:ARtarget})
\end{itemize}\end{varwidth}};
\node (tuning) [text width=3cm, left of=tuning_box, xshift = -7.5cm] {\LARGE \textbf{Tuning:}};
\node (InputTuning) [draw, ellipse, right of=tuning_box, xshift = 10.5cm] {\large \begin{varwidth}{\linewidth} $N_{\text{tune}}$ \\ $D$ \\ $\alpha_{\text{target}}$ \end{varwidth}};
\node (tuning) [text width=3cm, above of=InputTuning] {\LARGE Input:};
\node (burnin_box1) [draw, align=left, fill=white!30, below of=tuning_box, yshift = -0.7cm] {\large \begin{varwidth}{\linewidth}\begin{itemize}
\item[1.] Run $N_{\text{burn-in}}$ HMC iterations: VV, $L = 1$, $\Delta t_{\text{VV}}$
\item[2.] Store $N_{\text{acc}}$ and calculate AR \eqref{eq:ARcalculation} for $N = N_{\text{burn-in}}$
\end{itemize}\end{varwidth}};
\node (burnin) [text width=3cm, left of=burnin_box1, xshift = -7.5cm] {\LARGE \textbf{Burn-in:}};
\node (InputBurnin) [draw, ellipse, right of=burnin_box1, xshift = 10.5cm] {\large \begin{varwidth}{\linewidth} $N_{\text{burn-in}}$ \\ $I_{\omega} \in \{0, 1\}$ \end{varwidth}};
\node (ifIomega0) [draw, align=left, fill=white!30, below of=burnin_box1, yshift = -1cm, xshift = -3cm] {\large \textbf{Calculate}: \\ \large $\tilde{\omega} = \max \omega_j$ \eqref{eq:Frequencies} \\ \large $S$ \eqref{eq:FittingFactorsExplicit}
};
\node (ifSgreater2) [draw, align=left, fill=white!30, right of=ifIomega0, xshift = 3.75cm] {\large \textbf{Calculate}: \\ \large $\omega_j$ \eqref{eq:Frequencies}, $\sigma$ \\ \large $S_{\omega}$ \eqref{eq:FittingFactorsExplicit}};
\node (ifsigmaless1) [draw, align=left, fill=violet!30, below of=burnin_box1, yshift = -4cm] {\large \textbf{Calculate}: \\ \large $\text{SL}$ \eqref{ineq:StabilityLimitCalculation} \\ \large for $S_f = S_{\omega}$};
\node (ifSless2) [draw, align=left, fill=orange!40, left of=ifsigmaless1, xshift = -3cm] {\large \textbf{Calculate}: \\ \large $\text{SL}$ \eqref{ineq:StabilityLimitCalculation} \\ \large for $S_f = S$};
\node (ifsigmagreater1) [draw, align=left, fill=lime!50, right of=ifsigmaless1, xshift = 3cm] {\large \textbf{Calculate}: \\ \large $\text{SL}$ \eqref{ineq:StabilityLimitCalculationStdev}};
\node (randomizeSL) [draw, align=left, fill=white!30, below of=burnin_box1, yshift = -7cm] {\large \begin{varwidth}{\linewidth}\begin{itemize}
\item[1.] Compute $\mathfrak{R} (\text{SL})$ - a randomized stability limit
\item[2.] Produce a set of $N_{\text{pr}}$ randomized step sizes $\{\Delta t_{\text{pr}_i}\}_{i = 1, ... , N_{\text{pr}}}$ such that $\Delta t_{\text{pr}_i} = \mathfrak{R} (\Delta t_{\text{pr}}) \le \mathfrak{R} (\text{SL})$
\end{itemize}\end{varwidth}};
\node (SLRandomScheme) [draw, ellipse, align=left, fill=white!30, right of=randomizeSL, xshift = 10.5cm, yshift = 2.3cm] {\large \begin{varwidth}{\linewidth} randomization scheme  $\mathcal{R}$ \\ randomization interval $\mathcal{I}_{\mathcal{R}}$ \\ $\mathfrak{R} \equiv \{ \mathcal{R}, \mathcal{I}_{\mathcal{R}} \}$ \end{varwidth}};
\node (InputProduction) [draw, ellipse, align=left, fill=white!30, right of=randomizeSL, xshift = 10.5cm, yshift = -1.4cm] {\large $N_{\text{pr}}$};
\node (DeltatInput) [draw, ellipse, align=left, fill=white!30, right of=randomizeSL, xshift = 10.5cm] {\large $\Delta t_{\text{pr}}$};
\node (hSLwithSw) [draw, align=left, fill=violet!30, below of=randomizeSL, yshift = -1.2cm] {\large \textbf{Compute}: \\ $\{ h_{\text{pr}_i} \}_{i = 1, ... , N_{\text{pr}}}$ \eqref{eq:NondimensionalizationSfreqs}};
\node (hSLwithS) [draw, align=left, fill=orange!40, left of=hSLwithSw, xshift = -3cm] {\large \textbf{Compute}: \\ $\{ h_{\text{pr}_i} \}_{i = 1, ... , N_{\text{pr}}}$ \eqref{eq:NondimensionalizationSnofreqs}};
\node (hSLwithSwstdev) [draw, align=left, fill=lime!50, right of=hSLwithSw, xshift = 3cm] {\large \textbf{Compute}: \\ $\{ h_{\text{pr}_i} \}_{i = 1, ... , N_{\text{pr}}}$ \eqref{eq:NondimensionalizationSfreqsStdev}};
\node (bopt) [draw, align=left, fill=white, below of=hSLwithSw] {\large For each $h_{\text{pr}_i}$, $i = 1, ... , N_{\text{pr}}$ find $b_{{\text{opt}}_i}^k$ \eqref{eq:boptsAIA} \\ \large using the pretabulated map $\overline{h} \to b^k_{\text{opt}}$};
\node (production_box) [draw, align=left, fill=white, below of=bopt] {\large Run $N_{\text{pr}}$ HMC iterations using \eqref{eq:2stageIntegrators} ($k = 2$) or \eqref{eq:3stageIntegrators} ($k = 3$) \\ \large with $b_{{\text{opt}}_i}^k$ (for $k = 3$, $a_{\text{opt}_i}$ \eqref{3stageHyperbola}), $L_{\text{pr}_i}$, $\Delta t_{\text{pr}_i}$, $i = 1, ... , N_{\text{pr}}$};
\node (production) [text width=3cm, left of=production_box, xshift = -7.5cm] {\LARGE \textbf{Production:}};
\node (averages) [draw, align=left, fill=white, below of=production_box] {\large Compute averages from the generated HMC trajectory};
\draw [arrow] (tuning_box) -- (burnin_box1);
\draw [arrow] (InputTuning) -- (tuning_box);
\draw [arrow] (InputBurnin) -- (burnin_box1);
\draw [arrow] (burnin_box1) -- node[anchor=east] {\large $I_{\omega} = 0$} (ifIomega0);
\draw [arrow] (burnin_box1) -- node[anchor=west] {\large $I_{\omega} = 1$} (ifSgreater2);
\draw [arrow] (ifIomega0) -- node[anchor=south] {\large $S > 2$} (ifSgreater2);
\draw [arrow] (ifIomega0) -- node[anchor=east] {\large $S \le 2$} (ifSless2);
\draw [arrow] (ifSgreater2) -- node[anchor=east] {\large $\sigma \le 1$} (ifsigmaless1);
\draw [arrow] (ifSgreater2) -- node[anchor=west] {\large $\sigma > 1$} (ifsigmagreater1);
\draw [orange, arrows = {-Stealth[color=orange]}] (ifSless2) -- (randomizeSL);
\draw [violet!50, arrows = {-Stealth[color=violet!50]}] (ifsigmaless1) -- (randomizeSL);
\draw [lime, arrows = {-Stealth[color=lime]}] (ifsigmagreater1) -- (randomizeSL);
\draw [arrow] (SLRandomScheme) -- (randomizeSL.east);
\draw [arrow] (DeltatInput) -- (randomizeSL.east);
\draw [arrow] (InputProduction.west) -- (randomizeSL.east);
\draw [orange, arrows = {-Stealth[color=orange]}] (randomizeSL) -- (hSLwithS);
\draw [violet!50, arrows = {-Stealth[color=violet!50]}] (randomizeSL) -- (hSLwithSw);
\draw [lime, arrows = {-Stealth[color=lime]}] (randomizeSL) -- (hSLwithSwstdev);
\draw [orange, arrows = {-Stealth[color=orange]}] (hSLwithS) -- (bopt);
\draw [violet!50, arrows = {-Stealth[color=violet!50]}] (hSLwithSw) -- (bopt);
\draw [lime, arrows = {-Stealth[color=lime]}] (hSLwithSwstdev) -- (bopt);
\draw [arrow] (bopt) -- (production_box);
\draw [arrow] (production_box) -- (averages);
\draw [black!50, dashed] (8,-20.5) -- (8,1.2);
\end{tikzpicture}}}
\caption{Detailed schematic representation of the s-AIA$k$ algorithm.}\label{fig:saia}
\end{figure}

The s-AIA algorithm is summarized in Figure~\ref{fig:sAIA_chart}.
Given a model; a dataset; HMC parameters and settings for Tuning, Burn-in and Production stages; $I_{\omega}$ (see Figure~\ref{fig:sAIA_chart}) and an order $k$ of s-AIA ($k = 2$ or $3$), s-AIA algorithm works as shown in Figure~\ref{fig:saia}.
We remark that there are no particular requirements in the algorithm regarding a choice of randomization schemes for step sizes or trajectory lengths. Some of such schemes will be discussed in Section \ref{sec:NumericalResults}.

\begin{sloppypar}
s-AIA has been implemented in the BCAM in-house software package \textsf{HaiCS} (Hamiltonians in Computational Statistics) for statistical sampling of high dimensional and complex distributions and parameter estimation in Bayesian  models using MCMC and HMC based methods.
A detailed presentation and description of the package can be found in \cite{tijana_thesis}, whereas applications of \textsf{HaiCS} software are presented in \cite{radivojevic_akhmatskaya_MHMC_2020, radivojevic2018, inouzhe2023}.
\end{sloppypar}

\section{Numerical results and discussion}\label{sec:NumericalResults}
In order to evaluate the efficiency of the proposed s-AIA algorithms, we compared them in accuracy and performance with the integrators previously introduced for HMC-based sampling methods (Table \ref{tab:IntegratorsTable}). We examined  2- and 3-stage s-AIA on four benchmark models presented.

\subsection{Benchmarks}\label{sec:SimulationBenchmarks}
\begin{figure}[ht!]
\centering
\includegraphics[width = \textwidth]{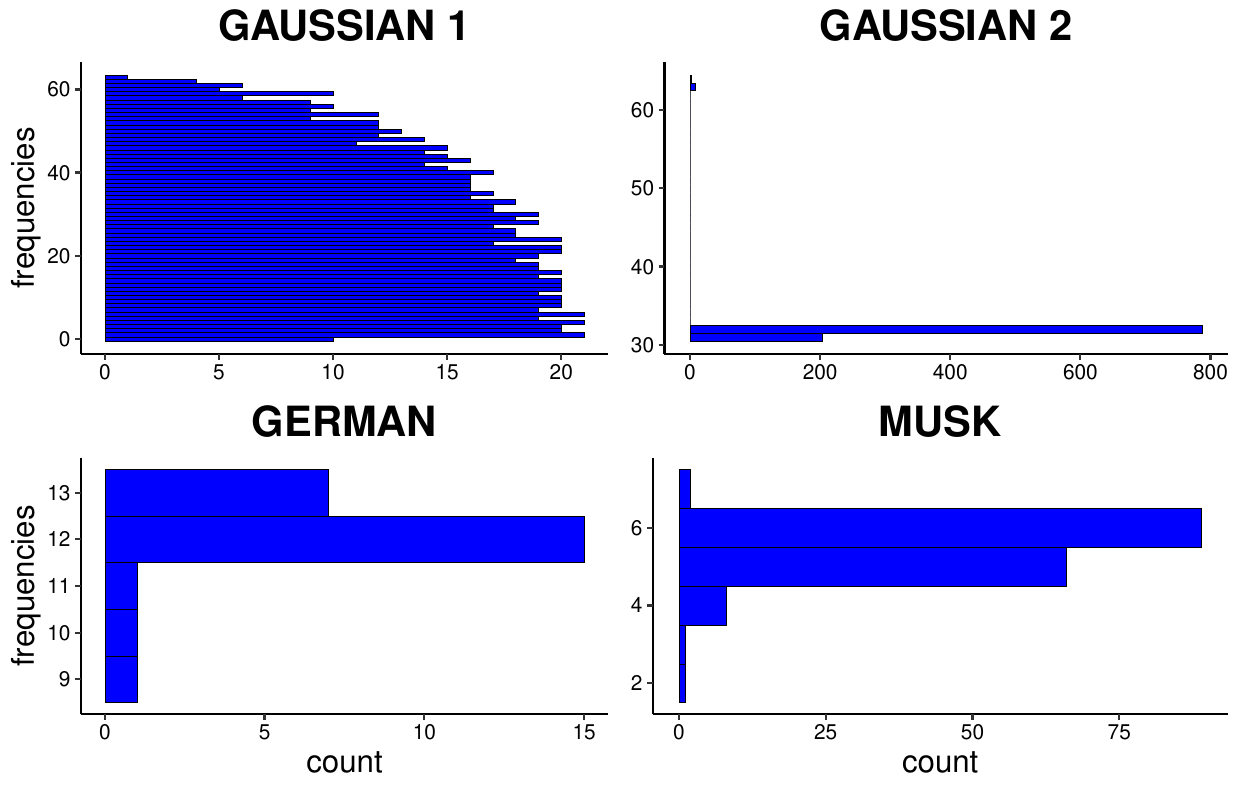}
 \caption{Frequency distributions of the benchmark models.}
 \label{fig:frequencies}
\end{figure}

\begin{sloppypar}
\begin{itemize}
\item \textbf{Gaussian 1}, \textbf{Gaussian 2}: two $D$-dimensional multivariate Gaussian models $\mathcal{N} (0, \Sigma)$, $D = 1000$, with precision matrix $\Sigma^{-1}$ generated from a Wishart distribution with $D$ degrees of freedom and the $D$-dimensional identity scale matrix \cite{hoffman_gelman2014} (Gaussian 1) and with diagonal precision matrix $\Sigma^{-1}$ made by $D_1 = 990$ elements taken from $\mathcal{N} (1000, 100)$ and $D_2 = 10$ from $\mathcal{N} (4000, 1600)$ (Gaussian 2).
\item \textbf{German}, \textbf{Musk}: two real datasets for a Bayesian Logistic Regression model \cite{liu2001, radivojevic_akhmatskaya_MHMC_2020} available from the University of California Irvine Machine Learning Repository \cite{lichman2013uci}, with dimensions $D = 25 \, \text{(German)}, 167 \, \text{(Musk)}$ and $K = 1000 \, \text{(German)}, 476 \, \text{(Musk)}$ observations.
\end{itemize}
The frequency distributions of the selected benchmarks estimated as proposed in Section \ref{sec:extensionAIAtoCompStat} are plotted in Figure~\ref{fig:frequencies}.
\end{sloppypar}

\subsection{Metrics}\label{sec:Metrics}
For HMC performance evaluation we monitored the following properties:
\begin{itemize}
\item \textbf{Acceptance rate.} The acceptance rate (AR) is the ratio between the accepted and the total $N$ number of proposals as in \eqref{eq:ARcalculation}.
\item \textbf{Effective Sample Size.}
The Effective Sample Size (ESS) is the number of effectively uncorrelated samples out of $N$ collected samples of a Markov chain. We calculated it, as proposed in \cite{radivojevic_akhmatskaya_MHMC_2020}, through the \textit{effectiveSize} function of the \textsf{CODA} package of \textsf{R} \cite{plummer2006coda}.
\item \textbf{Monte Carlo Standard Error.}
The Monte Carlo Standard Error (MCSE) quantifies the estimation noise caused by  Monte Carlo sampling methods. It indicates the estimated Standard Error of the sample mean
\begin{equation*}
\hat{\bm{\mu}} = \frac{1}{N} \sum_{i=1}^N \bmtheta_i
\end{equation*}
in a Markov chain \cite{kruschke2015}, and is calculated by substituting the sample size $N$ in the Standard Error formula
\begin{equation}\label{eq:StandardError}
\text{SE} = \sqrt{\frac{\hat{\bm{\sigma}}^2}{N}},
\end{equation}
with the ESS, i.e.
\begin{equation}\label{eq:MCSE}
\text{MCSE} = \sqrt{\frac{\hat{\bm{\sigma}}^2}{\text{ESS}}}.
\end{equation}
In \eqref{eq:StandardError} and \eqref{eq:MCSE}, $\hat{\bm{\sigma}}^2$ is an estimator of the sample variance \cite{vehtari2021}.
\item \textbf{Potential Scale Reduction Factor.}
The Potential Scale Reduction Factor (PSRF) monitors the convergence of a Markov chain by comparing it with other randomly initialized chains \cite{gelman_rubin1992}. We calculated it as explained in \cite{brooks_gelman1998} (Sections 1.2-1.3).
\end{itemize}

We took $\min \text{ESS}$ and $\min \left(\text{MCSE} \right)^{-1}$ normalized with respect to the theoretical average number of gradient evaluations, that is $k \bar{L}$ ($\bar{L}$ is the theoretical average of number of integration steps, $k$ is the number of stages of an integrator in use). Evaluation of gradients constitutes the bulk of the computational effort in HMC simulations and the chosen normalization leads to fair comparison between integrators with different number of stages. Of course, larger values of $\min \text{ESS}$ and $\min \left(\text{MCSE} \right)^{-1}$ imply better sampling performance.

Finally, we monitored $\max \text{PSRF}$ to examine the convergence of  tests and used a very conservative threshold, $\text{PSRF} < 1.01$, as suggested in \cite{vehtari2021}, for all benchmarks but Musk, for which the threshold was relaxed to $1.1$ \cite{gelman_rubin1992}. We remark that the popular approach for the ESS and MCSE calculations by Geyer \cite{geyer1992} (as implemented in Stan \cite{STAN}) was also tested with the proposed benchmarks and produced almost identical values. We omit those results for brevity.

\subsection{Simulation setup}\label{sec:SimulationSetup}
The proposed $k$-stage s-AIA algorithms, $k = 2, 3$, were tested for a range of step sizes $\{ k \Delta t_i \}$ within the system-specific dimensional stability interval $\left( 0, k \Delta t_{\text{SL}} \right)$. Such an interval is found through the dimensionalization of the theoretically predicted nondimensional stability limit for the $k$-stage Velocity Verlet using the fitting factor \eqref{eq:FittingFactorsExplicit} and a method chosen among \eqref{ineq:StabilityLimitCalculation}, \eqref{ineq:StabilityLimitCalculationStdev}.
To realize the randomization of each tested step size within the stability interval, the interval was adjusted to the heuristically chosen randomization scheme, i.e. it is increased or decreased by a benchmark-specific $k \delta t$ as detailed in Table \ref{tab:SimulationSetup}. Afterwards, we built a grid of step sizes $\{ k \Delta t_i \}_{i = 1, ... , 20}$ of the modified stability interval, where $k \Delta t_i = i \frac{k \Delta t_{\text{SL}}}{20}$, $i = 1, ..., 20$ and, for each $k \Delta t_i$, $i = 1, ..., 20$, we drew a step size for the simulation either from $\mathcal{U} (k \Delta t_i + k \delta t, k \Delta t_i)$, if $\delta t < 0$, or $\mathcal{U} (k \Delta t_i, k \Delta t_i + k \delta t)$, if $\delta t > 0$.
The number of integration steps per iteration, $L$, was drawn randomly uniformly at each iteration from  $\{1, ... , 2 \bar{L} - 1 \}$, with $\bar{L}$ such that
\begin{equation}\label{eq:TrajectoryLengthScheme}
\bar{L} h = \tau D,
\end{equation}
where $D$ is the problem dimension and $\tau$ is a benchmark-specific constant, found empirically to maximize performance near the center of the stability interval $h = k$. Such a setting provides a fair comparison between various multi-stage integrators by fixing the average number of gradients evaluations performed within each tested integrator. We remark that optimal choices of HMC simulation parameters, such as step sizes, numbers of integration steps and randomization intervals are beyond the scope of this study and will be discussed in detail  elsewhere.
Each simulation was repeated 10 times and the results reported in the paper were obtained by averaging over those multiple runs to reduce statistical errors. The simulation settings are detailed in Table \ref{tab:SimulationSetup}.

\begin{table}[h!]
\centering
\resizebox{\textwidth}{!}{
\begin{tabular}{c c c l c c c c}
Benchmark & $D$ & $N_{\text{pr}}$ & Fitting factor & $\sigma$ correction & $\Delta t_{\text{SL}}$ & $k \bar{L}$ & $\delta t$ \\
\hline
\multirow{2}{*}{Gaussian 1} & \multirow{2}{*}{$1000$} & \multirow{2}{*}{$20000$} & $S = 1$ & - & $0.03017$ & \multirow{2}{*}{$4000$} & \multirow{2}{*}{$-\frac{\Delta t_{\text{SL}}}{20}$} \\
& & & $S_{\omega} = 1.2648$ & yes ($\sigma = 16.7$) & $0.03248$ & & \\
\multirow{2}{*}{Gaussian 2} & \multirow{2}{*}{$1000$} & \multirow{2}{*}{$20000$} & $S = 1$ & - & $0.02983$ & \multirow{2}{*}{$1000$} & $-\frac{\Delta t_{\text{SL}}}{20}$ \\
& & & $S_{\omega} = 1.2641$ & yes ($\sigma = 3.14$) & $0.03005$ & & $\frac{3 \Delta t_{\text{SL}}}{20}$ \\
\multirow{2}{*}{German} & \multirow{2}{*}{$25$} & \multirow{2}{*}{$20000$} & $S = 1.3273$ & - & $0.1093$ & \multirow{2}{*}{$25$} & \multirow{2}{*}{$-\frac{\Delta t_{\text{SL}}}{20}$} \\
& & & $S_{\omega} = 1.4284$ & no ($\sigma = 0.897$) & $0.1015$ & & \\
\multirow{2}{*}{Musk} & \multirow{2}{*}{$167$} & \multirow{2}{*}{$100000$} & $S = 2.9719$ & - & $0.1030$ & \multirow{2}{*}{$167$} & \multirow{2}{*}{$-\frac{3 \Delta t_{\text{SL}}}{20}$} \\
& & & $S_{\omega} = 3.8827$ & no ($\sigma = 0.578$) & $0.07115$ & & \\
\hline
\end{tabular}
}
\caption{\label{tab:SimulationSetup} Parameters settings for each benchmark model: $D$ is the dimension of a benchmark, $N_{\text{pr}}$ is the number of iterations for the production stage (see Fig. \ref{fig:saia}), $\sigma$ is the standard deviation of the frequencies $\omega_j$ \eqref{eq:Frequencies}, $\Delta t_{\text{SL}}$ is the estimated stability limit, $k$ is the number of stages, $\bar{L}$ is the average number of integration steps per iteration \eqref{eq:TrajectoryLengthScheme}, $\delta t$ is the length of a randomization interval for the integration step size.}
\end{table}

\subsection{Results and discussion}
First, we tested 2- and 3-stage s-AIA integrators using the fitting factor approach $S_{\omega}$ \eqref{eq:FittingFactorsExplicit} and its corresponding nondimensionalization methods \eqref{eq:NondimensionalizationSfreqs} or \eqref{eq:NondimensionalizationSfreqsStdev}, selected according to the distribution of $\omega_j$ (Table \ref{tab:SimulationSetup}).

\begin{figure}[t!]
	\centering
	\includegraphics[width = \textwidth]{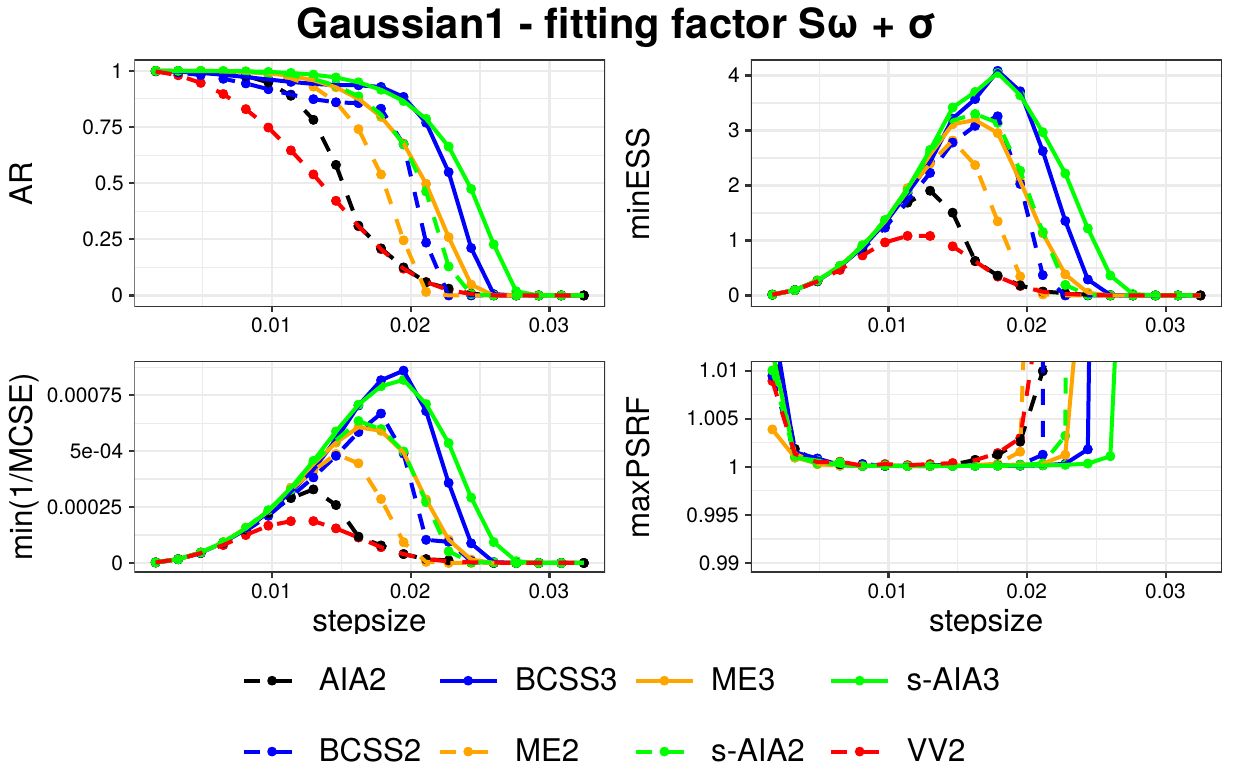}
	\caption{Gaussian 1 benchmark model with $S_{\omega}$ fitting factor \eqref{eq:FittingFactorsExplicit} and nondimensionalization \eqref{eq:NondimensionalizationSfreqsStdev}. The metrics in Section \ref{sec:Metrics} are plotted vs a range of step sizes within the stability  interval \eqref{ineq:StabilityLimitCalculationStdev}. s-AIA3 (solid green line) leads to the best HMC performance and improves on the other integrators for most step sizes and all the metrics. The $\max \text{PSRF}$ plot shows that s-AIA3 is the integrator with best convergence. s-AIA2 (dashed green line) shows similar advantages over the other 2-stage integration schemes. For both 2- and 3-stage s-AIA, the top performance in terms of $\min \text{ESS}$ and $\min \left({1}/{\text{MCSE}} \right)$ is reached near the center of the stability interval.}
	\label{fig:Gaussian1_S2stdev}
\end{figure}

\begin{figure}[ht!]
	\centering
	\includegraphics[width = \textwidth]{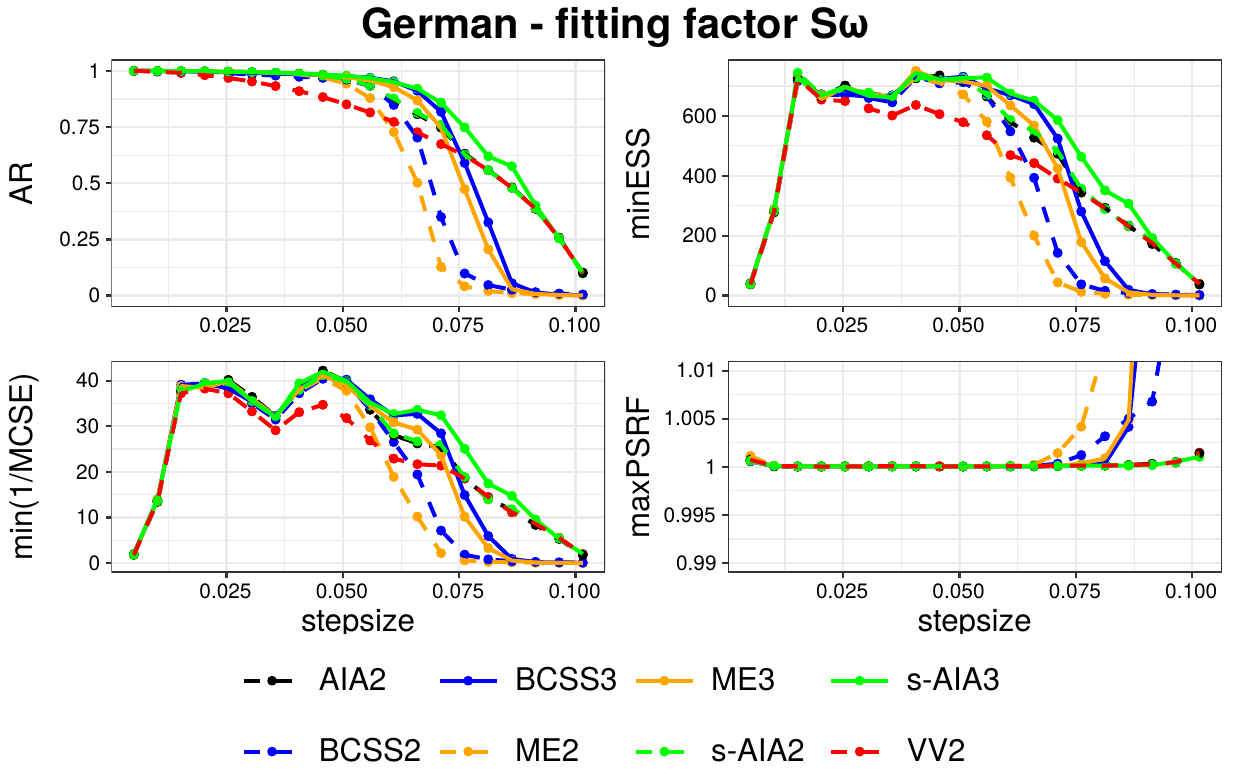}
	\caption{German benchmark model with $S_{\omega}$ fitting factor \eqref{eq:FittingFactorsExplicit} and nondimensionalization \eqref{eq:NondimensionalizationSfreqs}. The metrics in Section \ref{sec:Metrics} are plotted vs a range of step sizes within the stability  interval \eqref{ineq:StabilityLimitCalculation}. s-AIA3 (solid green line) improves on the other integrators for all step sizes and all the metrics. It shows its best performance near the center of the stability interval. s-AIA2 (dashed green line) shows similar advantages within the class of 2-stage integration schemes. The $\max \text{PSRF}$ plot shows that both s-AIA2 and s-AIA3, together with AIA2 (dashed black line) and VV2 (dashed red line), maintain convergence within the entire stability interval.}
	\label{fig:German_S2}
\end{figure}

Figures \ref{fig:Gaussian1_S2stdev}--\ref{fig:German_S2} show the  metrics collected for the Gaussian 1 and the German BLR benchmarks. One can appreciate the superiority of 2- and 3-stage s-AIA in terms of acceptance rate and sampling performance when compared with fixed-parameter multi-stage schemes of the same number  of stages. Recall that, as explained before, the standard Verlet typically used in HMC is included in the family of multi-stage schemes. In particular, s-AIA integrators reach the best possible performance in their groups, i.e. 2- and 3-stage groups respectively, almost for each step size in the stability interval. This means that the adaptation of the integrator coefficient $b^k_{\text{opt}}$ with respect to the randomized step size did enhance the accuracy and sampling of HMC. Specifically, the highest performance was reached around the center of the stability interval, in good agreement with the recommendations in \cite{bcss2014, mazur1997}. As expected, HMC combined with 3-stage s-AIA outperformed HMC with 2-stage s-AIA in sampling efficiency. Moreover, the $\max \text{PSRF}$ plot demonstrates that 3-stage s-AIA was the last integrator to lose convergence. In particular, for German BLR (Figure~\ref{fig:German_S2}), s-AIA ensured convergence over the entire range of step sizes, which suggests that the stability limit had been estimated accurately, i.e. the chosen fitting factor approach worked properly.

\begin{figure}[ht!]
	\centering
	\includegraphics[width = \textwidth]{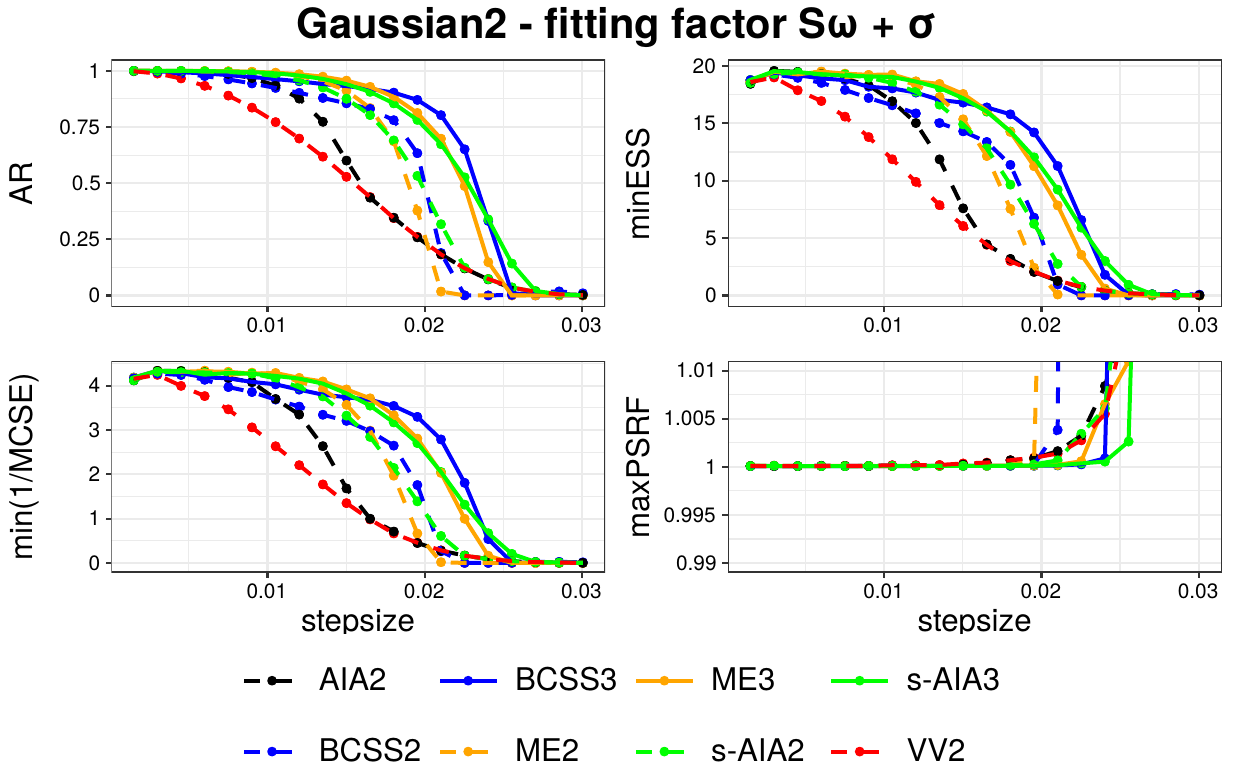}
	\caption{Gaussian 2 benchmark model with $S_{\omega}$ fitting factor \eqref{eq:FittingFactorsExplicit} and nondimensionalization \eqref{eq:NondimensionalizationSfreqsStdev}. Metrics are plotted vs a range of step sizes within the stability interval \eqref{ineq:StabilityLimitCalculationStdev}. s-AIA3 (solid green line) leads to performance comparable to the highest one for most step sizes. The $\max \text{PSRF}$ plot confirms that s-AIA3 guarantees the best HMC convergence. s-AIA2 (dashed green line) shows similar advantages within the class of 2-stage integration schemes.}
	\label{fig:Gaussian2_S2stdev}
\end{figure}

\begin{figure}[ht!]
	\centering
	\includegraphics[width = \textwidth]{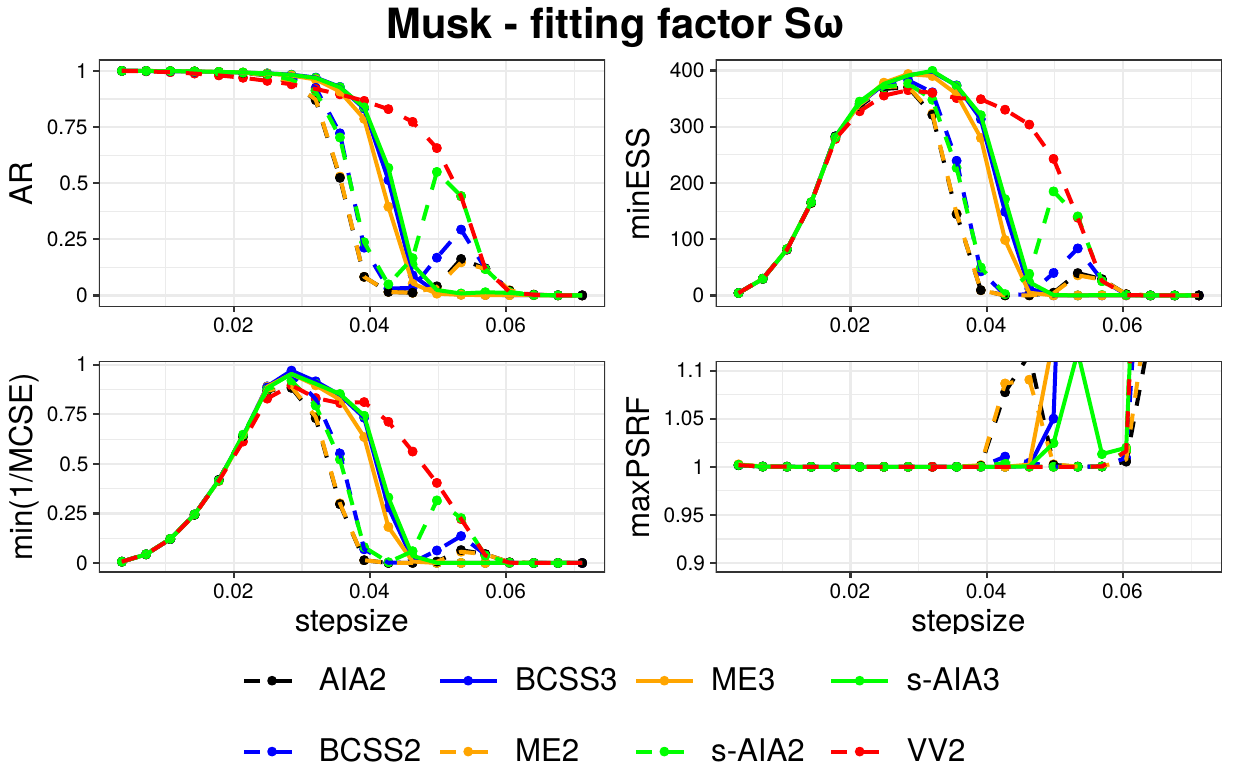}
	\caption{Musk with $S_{\omega}$ fitting factor \eqref{eq:FittingFactorsExplicit} and nondimensionalization  \eqref{eq:NondimensionalizationSfreqs}. Metrics are plotted vs a range of step sizes within the stability interval \eqref{ineq:StabilityLimitCalculation}. s-AIA3 (solid green) leads to the best performance together with ME3 (solid orange) and BCSS3 (solid blue), while VV2 (dashed red) maintains better performance for larger step sizes. The AR and $\max \text{PSRF}$ plots indicate that the stability  interval is overestimated.}
	\label{fig:Musk_S2}
\end{figure}

\begin{figure}[ht!]
	\centering
	\includegraphics[width = \textwidth]{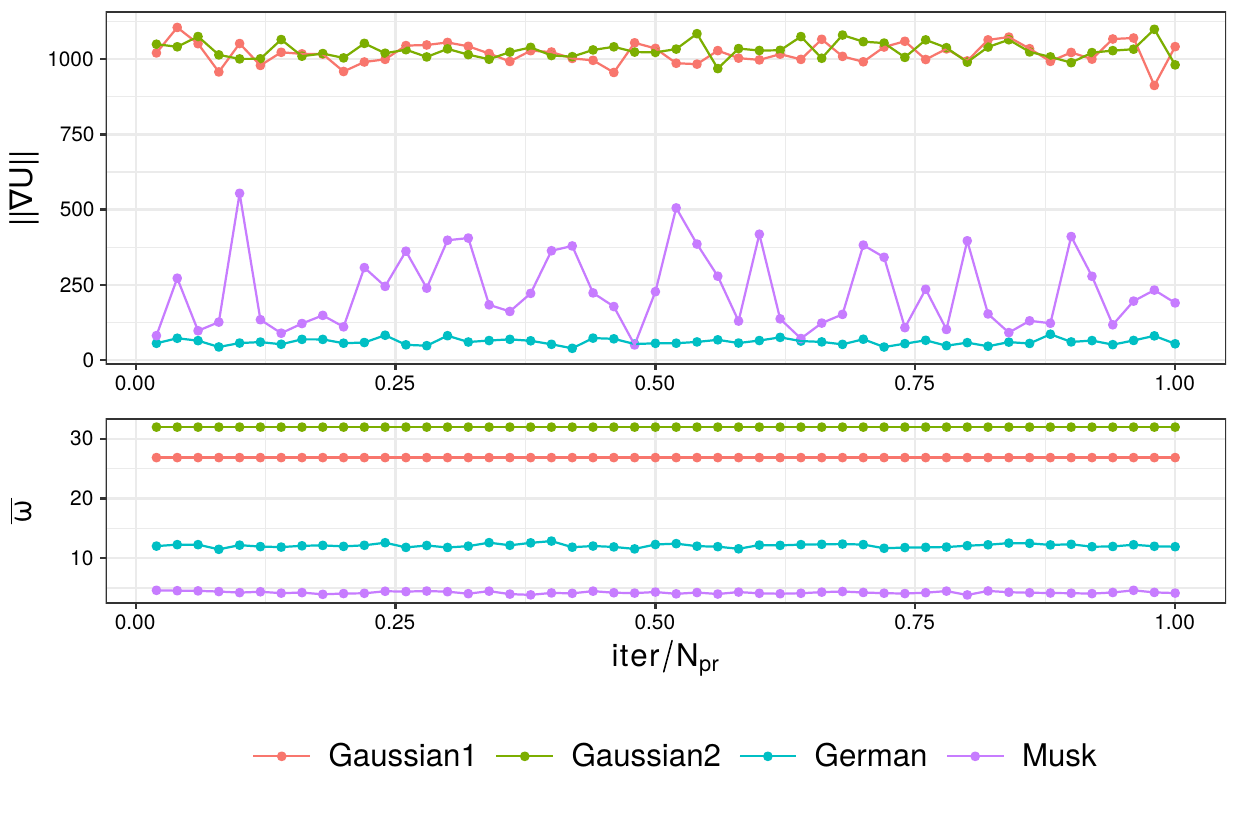}
	\caption{Evolution of the Euclidean norm of the forces $||\nabla U||$ (top) and average frequencies $\bar{\omega}$ (bottom) observed in the numerical experiments for the four benchmarks with $S_{\omega}$ fitting factor (Figures \ref{fig:Gaussian1_S2stdev}--\ref{fig:Musk_S2}). On the horizontal axis: $N_{\text{pr}}$ is the total number of iterations during the production stage (see Table \ref{tab:SimulationSetup}), $iter = \frac{i}{50} N_{\text{pr}}$, $i = 1, ... , 50$. The low frequencies of Musk BLR (violet line, bottom plot) generate the anharmonic behavior (violet line, top plot).}
	\label{fig:gradients_frequencies}
\end{figure}

Similar trends, though  less pronounced, can be observed for the Gaussian 2 benchmark in Figure~\ref{fig:Gaussian2_S2stdev}.
The top results achieved by 2- and 3-stage s-AIA are comparable to those demonstrated by the best behaved for this benchmark BCSS and ME integrators. Again, 3-stage s-AIA showed clear superiority over its 2-stage counterparts, and both turned out to be the last integrators to lose convergence in their groups.
In contrast, the same fitting factor approach $S_{\omega}$ applied to the Musk BLR benchmark did not show the level of accuracy observed for other benchmarks. In Figure~\ref{fig:Musk_S2}, one can admit the poor performance achieved for almost all integrators in the second half of the stability  interval, i.e. the stability limit was overestimated. However, 3-stage s-AIA reached the best values in terms of $\min \text{ESS}$ and $\min \left(\text{MCSE} \right)^{-1}$, again around the center of the stability interval.
Further analysis of the simulated frequencies and forces of the benchmarks revealed (see Figure~\ref{fig:gradients_frequencies}) the anharmonic behavior of the Musk system, which, along with the fitting factor $S \approx 2.93 > 2$ (Table \ref{tab:SimulationSetup}), explains the inaccuracy of the harmonic analysis presented in Section \ref{sec:extensionAIAtoCompStat} (\emph{Calculation of fitting factors}) in the estimation of the stability limit in this case.

\begin{figure}[ht!]
	\centering
	\includegraphics[width = \textwidth]{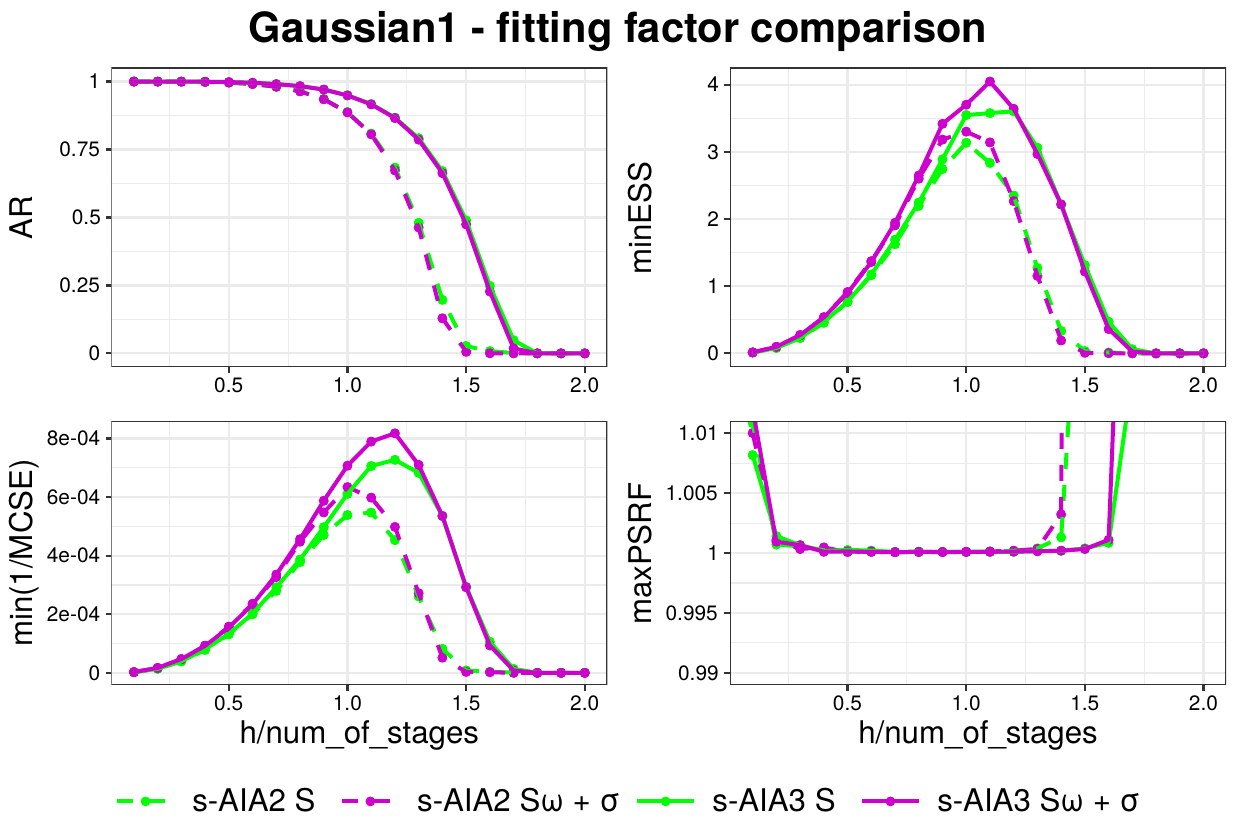}
	\caption{Gaussian 1: The effect on the HMC performance of different scaling approaches $S$ and $S_{\omega}$ \eqref{eq:FittingFactorsExplicit} with \eqref{eq:NondimensionalizationSnofreqs} (in green) and with \eqref{eq:NondimensionalizationSfreqsStdev} (in purple) respectively. The metrics to monitor are plotted against the 1-stage dimensionless stability  interval $(0, 2)$ in order to display the comparison. HMC with s-AIA using the $S_{\omega}$ fitting factor approach (in purple) exhibits more accuracy and better sampling around the center of the stability interval.}
	\label{fig:Gaussian1_comparison}
\end{figure}

\begin{figure}[h!]
	\centering
	\includegraphics[width = \textwidth]{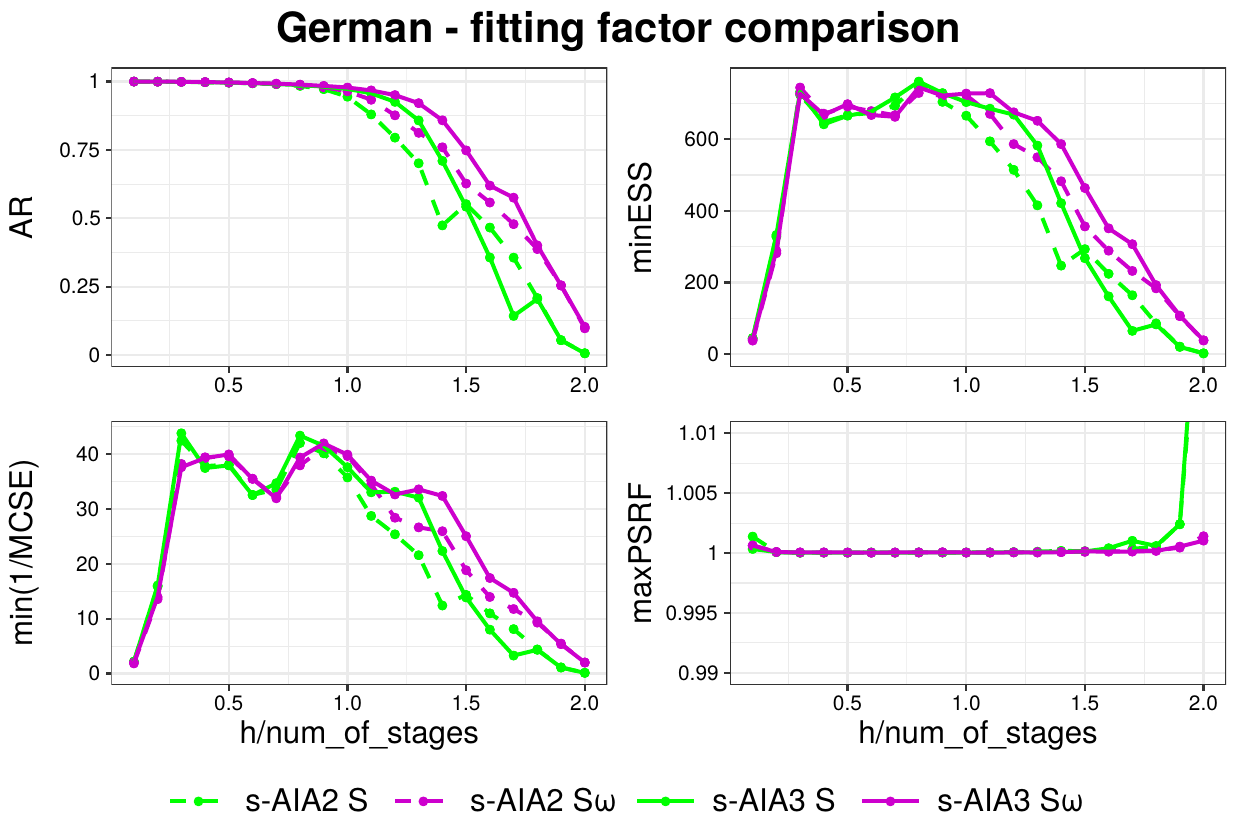}
	\caption{German BLR: The effect on the HMC performance of different scaling approaches $S$ and $S_{\omega}$ \eqref{eq:FittingFactorsExplicit} with \eqref{eq:NondimensionalizationSnofreqs} (in green) and with \eqref{eq:NondimensionalizationSfreqs} (in purple) respectively. Metrics are plotted against the 1-stage dimensionless stability  interval $(0, 2)$ for comparison. HMC with s-AIA using the $S_{\omega}$ fitting factor (in purple) exhibits more accuracy and better sampling in the second part of the stability interval.}
	\label{fig:German_comparison}
\end{figure}

\begin{figure}[h!]
	\centering
	\includegraphics[width = \textwidth]{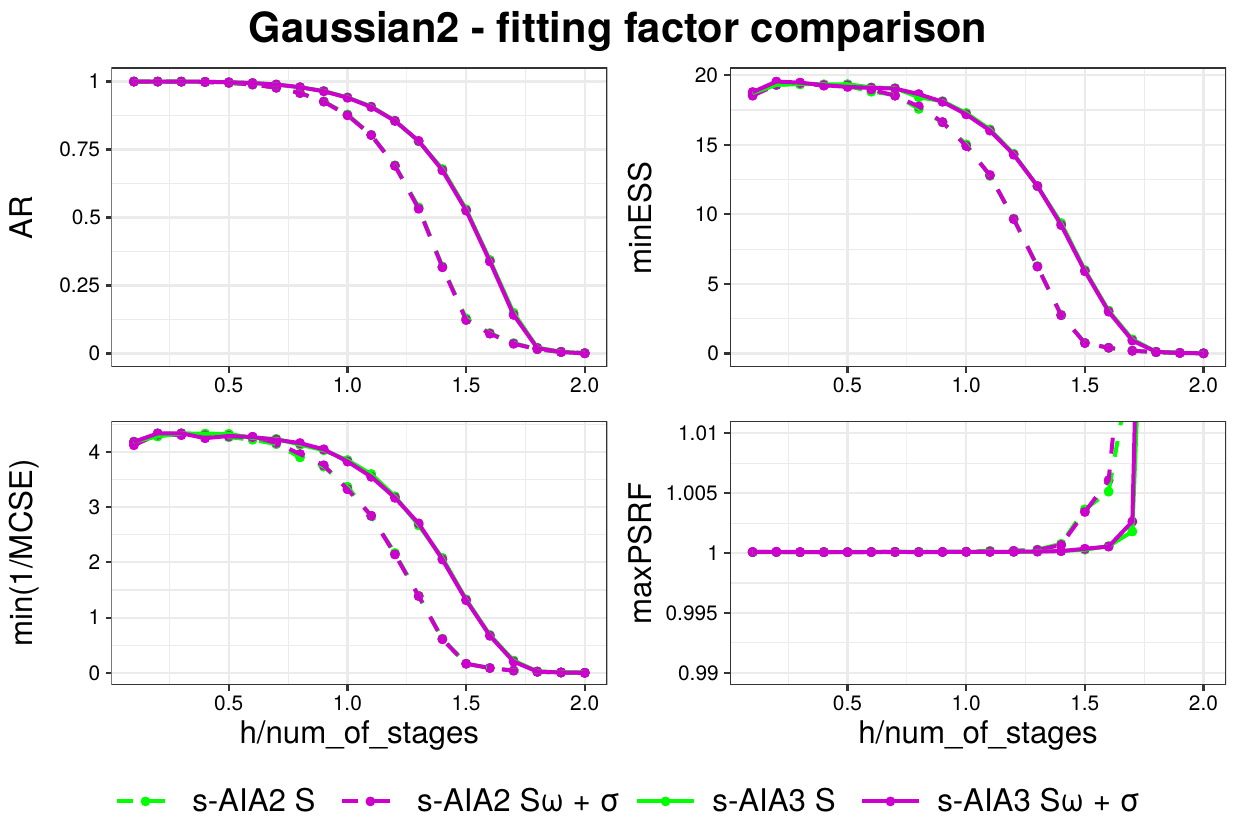}
	\caption{Gaussian 2: The effect on the HMC performance of the scaling approaches $S$ and $S_{\omega}$ \eqref{eq:FittingFactorsExplicit} with \eqref{eq:NondimensionalizationSnofreqs} (in green) and with \eqref{eq:NondimensionalizationSfreqsStdev} (in purple) respectively. Metrics to monitor are plotted against the 1-stage dimensionless stability interval $(0, 2)$ for comparison. Both approaches lead to almost identical performance in terms of accuracy, sampling and stability.}
	\label{fig:Gaussian2_comparison}
\end{figure}

\begin{figure}[h!]
	\centering
	\includegraphics[width = \textwidth]{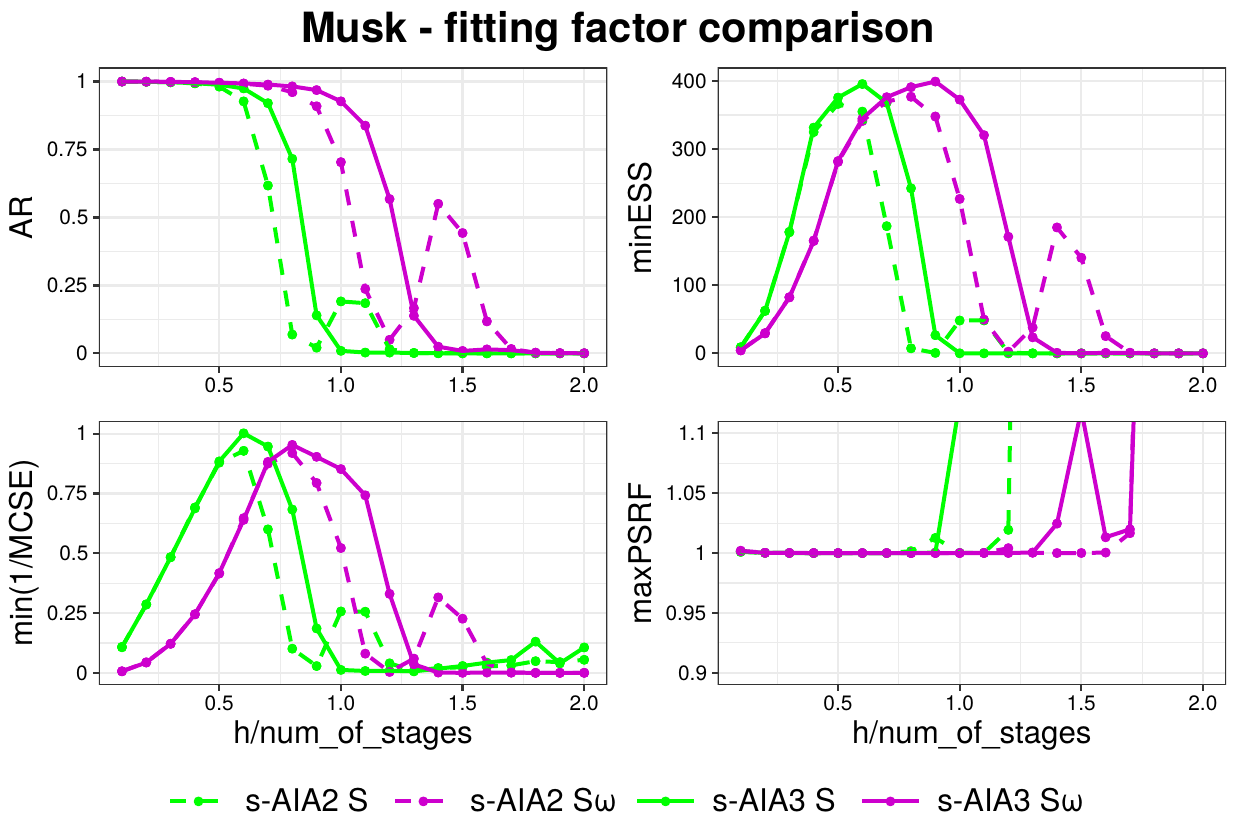}
	\caption{Musk BLR: The effect on the HMC performance of the scaling approaches $S$ and $S_{\omega}$ \eqref{eq:FittingFactorsExplicit} with \eqref{eq:NondimensionalizationSnofreqs} (in green) and with \eqref{eq:NondimensionalizationSfreqs} (in purple) respectively. Metrics are plotted against the 1-stage dimensionless stability interval $(0, 2)$ for comparison. Using $S_{\omega}$ (in purple) helps to shift the best performance of both adaptive schemes towards the center of the stability interval. Moreover, AR and $\max \text{PSRF}$ confirm that the stability limit is estimated better with $S_{\omega}$.}
	\label{fig:Musk_comparison}
\end{figure}

Next, we tested 2- and 3-stage s-AIA integrators using the fitting factor approach $S$ \eqref{eq:FittingFactorsExplicit} and its corresponding nondimensionalization method \eqref{eq:NondimensionalizationSnofreqs} (see Figures~\ref{fig:Gaussian1_comparison}--\ref{fig:Musk_comparison}).
As expected, the more accurate $S_{\omega}$ fitting factor and its nondimensionalization methods \eqref{eq:NondimensionalizationSfreqs}, \eqref{eq:NondimensionalizationSfreqsStdev} lead to an overall better performance than s-AIA with $S$ and \eqref{eq:NondimensionalizationSnofreqs}. However, for models with $S < 2$ (cf. Figures~\ref{fig:Gaussian1_comparison}, \ref{fig:German_comparison}, \ref{fig:Gaussian2_comparison}), both fitting approaches exhibited similar trends. On the other hand, for the Musk BLR benchmark, i.e. when $S > 2$, 2- and 3-stage s-AIA benefit from the more accurate $S_{\omega}$ fitting factor approach, reaching a clearly better estimation of the stability limit (cf. Figure~\ref{fig:Musk_comparison}).

Finally, we wish to review the behavior of the other tested multi-stage integrators. First, we remark the superiority of 3-stage integrators over their 2-stage counterparts. For any benchmark and fitting factor approach, the 3-stage integrators performed on average better at the same computational cost, as previously suggested in \cite{radivojevic2018}. In addition, we highlight that the other  integration schemes tested showed a strong dependence on the model in use. In particular, VV performed poorly for the Gaussian benchmarks (Figures~\ref{fig:Gaussian1_S2stdev}, \ref{fig:Gaussian2_S2stdev}) but demonstrated solid performance for the BLR models, especially for larger step sizes (Figures~\ref{fig:German_S2}, \ref{fig:Musk_S2}). Similarly to VV, AIA resulted to be one of the worst integrators for the Gaussian benchmarks (Figures~\ref{fig:Gaussian1_S2stdev}, \ref{fig:Gaussian2_S2stdev}), but achieved performance similar to 2-stage s-AIA for the BLR models (Figures~\ref{fig:German_S2}, \ref{fig:Musk_S2}). On the contrary, the BCSS and ME integrators performed similarly to s-AIA for Gaussian 2 and Musk (Figures~\ref{fig:Gaussian2_S2stdev}, \ref{fig:Musk_S2}), whereas they lose performance for Gaussian 1 and BLR German (Figures~\ref{fig:Gaussian1_S2stdev}, \ref{fig:German_S2}).

\begin{figure}[h!]
	\centering
	\includegraphics[width = \textwidth]{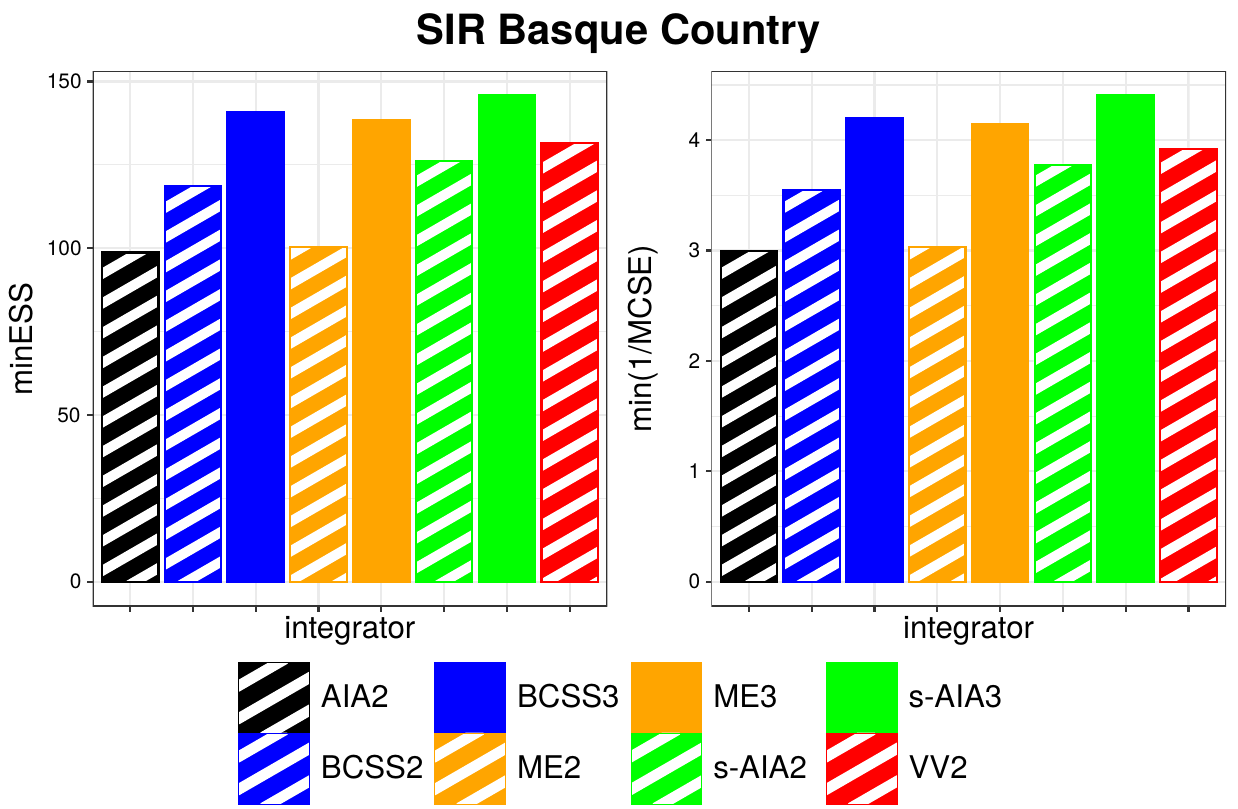}
	\caption{Standard SIR model combined with the COVID-19 daily incidence data from the Basque Country. Comparison of the minESS and $\text{min(MCSE)}^{-1}$ collected from HMC simulations with the integrators presented in Table \ref{tab:IntegratorsTable}. s-AIA3 (solid green bar) reaches the top performance with both performance metrics. 3-stage integrators (solid bars) outperform 2-stage counterparts (dashed bars).}
	\label{fig:SIR_standard}
\end{figure}

In addition to the benchmarks presented in \ref{sec:SimulationBenchmarks}, in order to test the efficiency of the s-AIA integrators on a more complex distribution, we considered a standard epidemiological model SIR \cite{SIR_paper} applied to the study of the transmission dynamics of COVID-19 in the Basque Country for the period from the 10th February 2020 to the 31st January 2021 \cite{inouzhe2023}.
The proposed model comprises systems of ODEs (see for details \ref{app:SIR}) to be solved at each HMC iteration, which implies time-consuming simulations.  For that reason, only one step size was considered for the model, namely the center of the estimated stability interval $\frac{k \Delta t_{\text{SL}}}{2}$, $k = 2, 3$. Such a choice is supported by our numerical experiments presented in this section.  
We tested both 2- and 3-stage s-AIA and compared their performance with those obtained using the integrators summarized in Table \ref{tab:IntegratorsTable}. The simulation step sizes were randomized, following the procedure described in Section \ref{sec:SimulationSetup}. Each simulation was repeated 10 times and the results reported in Figure \ref{fig:SIR_standard} were obtained by averaging over multiple runs to reduce statistical errors. The simulation parameters are detailed in Table \ref{tab:SIRSetup}. Due to the complexity of the model and significant computational costs involved, for the estimation of the stability interval, we followed the strategy proposed in Section \ref{sec:extensionAIAtoCompStat} (\emph{Calculation of fitting factors}), Eqs. \eqref{eq:FittingFactor}-\eqref{eq:Somegatilde}, \eqref{ineq:StabilityLimitCalculation}. We remark that the choice of the simulation length $N_{\text{pr}}$ was dictated by the complexity of the model, available computational resources and the illustrative purposes of the simulations. For real applications, longer simulations are recommended for achieving reliable results. 

\begin{table}[ht!]
\centering
\begin{tabular}{c c c l c c c c}
Model & $D$ & $N_{\text{pr}}$ & Fitting factor & $\Delta t_{\text{SL}}$ & $k \bar{L}$ & $\delta t$ \\
\hline
SIR & $4$ & $160000$ & $S \tilde{\omega} = 387.13$ & $0.0002583$ & $4$ & $\frac{\Delta t_{\text{SL}}}{10}$ \\
\hline
\end{tabular}
\caption{\label{tab:SIRSetup} Parameters settings for the SIR model: $D$ is the dimension of a benchmark, $N_{\text{pr}}$ is the number of iterations for the production stage (see Fig. \ref{fig:saia}), $\Delta t_{\text{SL}}$ is the estimated stability limit, $k$ is the number of stages, $\bar{L}$ is the average number of integration steps per iteration \eqref{eq:TrajectoryLengthScheme}, $\delta t$ is the length of a randomization interval for the integration step size.}
\end{table}

Figure \ref{fig:SIR_standard} confirms the superiority of the 3-stage s-AIA algorithm in terms of minESS and $\text{min(MCSE)}^{-1}$ for the standard SIR model. Moreover, as expected, 3-stage integrators outperform the 2-stage counterparts, whereas the 2-stage s-AIA along with the 2-stage Velocity Verlet demonstrate the best performance within their group. The similar behaviors of s-AIA3 and BCSS3 around the center of the stability interval suggest the accurate estimation of the stability limit.   

In conclusion, we observed that the s-AIA algorithms enhanced the performance of HMC, if the stability interval length was estimated accurately. When that is the case, s-AIA demonstrates the best performance around the center of the stability interval, which, together with \eqref{ineq:StabilityLimitCalculation}-\eqref{ineq:StabilityLimitCalculationStdev}, gives a helpful suggestion for the choice of  step size in  HMC simulations. Moreover, the more accurate fitting factor approach $S_{\omega}$ \eqref{eq:FittingFactorsExplicit} with \eqref{eq:NondimensionalizationSfreqs} or \eqref{eq:NondimensionalizationSfreqsStdev} provided a better approximation of the stability limit, which resulted in  higher accuracy and greater performance of the adaptive integrators, mostly when applied to systems with prevailing anharmonic forces, i.e. if $S > 2$.

\section{Conclusion}\label{sec:Conclusion}
We have presented a novel adaptive multi-stage integration approach for enhancing the accuracy and sampling efficiency of HMC-based methods for Bayesian inference applications. The proposed methodology, which we call s-AIA, provides, for any choice of  step size within the stability  interval, a system-specific palindromic 2- or 3-stage splitting integrator which ensures the best energy conservation for harmonic forces within its family. Moreover, we offered a solution for detecting a system specific dimensional stability  interval using the simulation data generated at the HMC burn-in stage. In particular, we introduced three optional scaling/nondimensionalization approaches for estimating the stability limit with different level of accuracy and computational effort.

s-AIA was implemented (without introducing computational overheads in simulations) in the in-house software package \textsf{HaiCS} (Hamiltonians in Computational Statistics) \cite{tijana_thesis, radivojevic_akhmatskaya_MHMC_2020} and tested against the popular numerical integrators (Verlet \cite{verlet1967, swope1982}, BCSS \cite{bcss2014} and Minimum Energy \cite{mclachlan1995, predescu2012}) on the range of benchmark models.
We found that the adaptivity helped to reach the best possible performance within the families of 2-, 3-stage splitting integration schemes. We emphasize that standard Velocity Verlet, the HMC integrator of choice, is a member of those families.
If the stability limit was estimated accurately, s-AIA integrators reached the best performance in their groups, i.e. 2- and 3-stage groups, almost for each step size in the stability  interval. Also, using more stages enhanced the sampling performance, stability and conservation of the energy of the harmonic forces with the same computational effort.

We have demonstrated that the more accurate fitting factor approach $S_{\omega}$ \eqref{eq:FittingFactorsExplicit} led to an overall better performance in HMC simulations than its less computationally expensive counterpart $S$. However, the latter was able to reach comparable results when lying below the upper threshold $S < 2$ \cite{schlick_etal1998}. In that way, computational time and resources may be saved by avoiding the computation of angular frequencies. On the other hand, for more complex distributions, e.g. with dominating low-frequencies (like the Musk BLR benchmark model \cite{lichman2013uci}), we found that a proper analysis of the underlying dynamics of the simulated system might assist in the choice of a suitable system-specific fitting factor, the randomization interval and the number of HMC iterations required for a chain to converge.

We remark that even in the case of a rough estimation of the stability limit (like in Musk BLR), HMC with multi-stage adaptive splitting schemes achieves top performance in comparison with the fixed-parameter schemes, though the exact location of the optimal step size is harder to predict in this case. In an upcoming study, we will show how the proposed methodology can be adjusted for refining optimal parameters of HMC-based simulations.

\appendix
\section{Derivation of \texorpdfstring{$\rho_3 (h, b)$}{} in \eqref{eq:rho3stage}}
\label{app:rho3stage}
Consider the harmonic oscillator with Hamiltonian
\begin{equation}\label{HamiltonianHarmonicOscillator}
H = \frac{1}{2} (p^2 + \theta^2), \qquad \theta,p \in \mathbb{R},
\end{equation}
and equations of motions
\begin{equation}\label{EqsHarmonicOscillator}
\frac{d \theta}{dt} = p, \qquad \frac{d p}{dt} = - \theta.
\end{equation}
Given a $k$-stage palindromic splitting integrator $\Psi_h$ ($h$ is the integration step size), it acts on a configuration $(\theta_i, p_i)$ at the $i$-th iteration as
\begin{equation}\label{NumericalIntegratorEqsofmotion}
\Psi_h \left(
\begin{array}{c}
q_i \\ p_i
\end{array}
\right) =
\left( \begin{array}{c}
q_{i+1} \\ p_{i+1}
\end{array} \right)
= \left( \begin{matrix}
A^{\bmz}_h & B^{\bmz}_h \\
C^{\bmz}_h & D^{\bmz}_h
\end{matrix} \right)
\left( \begin{array}{c}
q_i \\
p_i
\end{array} \right),
\end{equation}
for suitable method-dependent coefficients $A^{\bmz}_h$, $B^{\bmz}_h$, $C^{\bmz}_h$, $D^{\bmz}_h$ ($\bmz = \{b_i, a_j \}$ is the set of $k-1$ integration coefficients). In \cite{bcss2014}, a formula for $\rho (h, \bmz)$ is provided:
\begin{equation}\label{eq:rhoDefinition}
\rho (h, \bmz) = \frac{(B^{\bmz}_h + C^{\bmz}_h)^2}{2(1- {A^{\bmz}_h}^2)} .
\end{equation}
For a 3-stage palindromic splitting integrator \eqref{eq:3stageIntegrators}, the integrator coefficients are ($\bmz = \{b, a \}$)
\begin{align}
A^{\bmz}_h &= 1 - \frac{h^2}{2} + a (1/2 - b) (1/2 - a + b) h^4 - 2 a^2 b (1/2 - a) (1/2 - b)^2 h^6, \label{eq:A_h3stage} \\
B^{\bmz}_h &= h - 2 a (1 - a) (1/2 - b) h ^3 + 2 a^2 (1/2 - a) (1/2 - b)^2 h^5, \label{eq:B_h3stage} \\
C^{\bmz}_h &= - h + (2 a b (1 - b) - a/2 + 1/4) h^3 + \nonumber \\
& + 2 a b (1/2 - b) (a (1 - b) - 1/2) h^5 + 2 a^2 b^2 (1/2 - a) (1/2 - b)^2 h^7. \label{eq:C_h3stage}
\end{align}
Finally, for $a, b$ in \eqref{3stageHyperbola} and $A^{\bmz}_h$, $B^{\bmz}_h$ and $C^{\bmz}_h$ in \eqref{eq:A_h3stage}-\eqref{eq:B_h3stage}-\eqref{eq:C_h3stage}, $\rho (h, \bmz)$ in \eqref{eq:rhoDefinition} becomes
\begin{equation*}
\rho_3 (h, b) =
\scalebox{1}{$ \frac{h^4 \left(-3 b^4 + 8 b^3 -19/4 b^2 + b + b^2 h^2 \left(b^3 - 5/4 b^2 + b/2 - 1/16 \right) - 1/16 \right)^2}{2 \left(3 b - b h^2 \left(b - 1/4 \right) - 1 \right) \left(1 - 3 b - b h^2 \left(b - 1/2 \right)^2 \right) \left( -9 b^2 + 6 b - h^2 \left( b^3 - 5/4 b^2 + b/2 - 1/16 \right) - 1 \right)}$}.
\end{equation*}

\section{Derivation of \texorpdfstring{$\alpha_{\text{target}}$}{} for s-AIA tuning.}
For the burn-in stage, we use the 1-stage Velocity Verlet integrator with $L = 1$ and  step size $\Delta t_{\text{VV}}$, which should be ideally chosen to be close to the center of the stability interval to achieve the best accuracy and sampling efficiency of an HMC simulation \cite{mazur1997}. In order to identify such a step size, we estimate the expected acceptance probability $\mathbb{E} [\alpha]$ following \cite{calvo2021hmc} (Sec. 5.2, Th. 1), i.e.
\begin{equation}\label{eq:AtanFormula}
\mathbb{E} [\alpha] = 1 - \frac{2}{\pi} \arctan \sqrt{\frac{\mathbb{E} [\Delta H]}{2}} ,
\end{equation}
which holds for standard univariate Gaussian distribution, i.e. the harmonic oscillator with the Hamiltonian \eqref{HamiltonianHarmonicOscillator}, regardless of the integrator being used, the step size and $L$. For the burn-in stage simulation setting, the expected energy error $\mathbb{E} [\Delta H]$ is defined in \eqref{eq:EdH1sVV} and, evaluated at the middle of the stability interval, $h = 1$, i.e.
\begin{equation} \label{eq:E1VV_hsl}
\mathbb{E} [\Delta H] = \frac{1}{32}.
\end{equation}
Combining \eqref{eq:AtanFormula} and \eqref{eq:E1VV_hsl}, one obtains
\begin{equation*}
\mathbb{E} [\alpha] \approx 0.92 = \alpha_{\text{target}}.
\end{equation*}
We provide a detailed procedure for adjusting a step size $\Delta t_{\text{VV}}$ to reach $\alpha_{\text{target}}$ in Algorithm \ref{alg:TUNE}
\label{app:ARtarget}
\begin{algorithm}[!t]
\hrulefill
\begin{algorithmic}[1]
\Input number of iterations $N_{\text{tune}}$, dimension of the simulated system $D$, number of iterations for AR check $N_{\text{check}}$, target $\alpha_{\text{target}}$, sensibility $\epsilon > 0$, step size increment $\delta t > 0$
\Initialize $\Delta t_{\text{VV}} = \frac{1}{D}$, $N = N_{\text{tot}} = N_{\text{acc}} = 0$
\While {$N_{\text{tot}} + N_{\text{check}} < N_{\text{tune}}$}
\State Perform $N_{\text{check}}$ HMC iterations with VV and $L = 1$ \label{alg_line:HMCUpdate}
\State $N = N + N_{\text{check}}$
\State $N_{\text{acc}}$ number of acceptances over the last $N$ iterations
\State Compute AR \eqref{eq:ARcalculation}
\If{$\text{AR} < \alpha_{\text{target}} - \epsilon$}
\State $\Delta t_{\text{VV}} = \Delta t_{\text{VV}} - \delta t$
\State $N = 0$
\ElsIf{$\text{AR} > \alpha_{\text{target}} + \epsilon$}
\State $\Delta t_{\text{VV}} = \Delta t_{\text{VV}} + \delta t$
\State $N = 0$
\EndIf
\State $N_{\text{tot}} = N_{\text{tot}} + N_{\text{check}}$
\EndWhile
\Output $\Delta t_{\text{VV}}$
\end{algorithmic}
\hrulefill
\caption{Routine for tuning a step size $\Delta t_{\text{VV}}$ in a burn-in HMC simulation.}\label{alg:TUNE}
\end{algorithm}

\section{Derivation of \texorpdfstring{$\mathbb{E}^1_{\text{VV}}[\Delta H]$}{} in \eqref{eq:UnivariateEnergyErrorExpectation}}
\label{app:UnivEnergyErrorExpectation1sVV}
According to \cite{bcss2014}, for the harmonic oscillator with the Hamiltonian \eqref{HamiltonianHarmonicOscillator} and the equations of motion \eqref{EqsHarmonicOscillator}, the expected energy error produced by a $k$-stage palindromic splitting integrator $\Psi_h$ applied for $L$ integration steps is given by
\begin{equation}\label{eq:EdHAppendix}
\mathbb{E}[\Delta H] = \sin^2 \left( L \Theta^{\bmz}_h \right) \rho (h, \bmz),
\end{equation}
where $\Theta^{\bmz}_h = \arccos A^{\bmz}_h$, and $A^{\bmz}_h$ is defined in \eqref{NumericalIntegratorEqsofmotion}. For $L = 1$ and $\rho (h, \bmz)$ defined in \eqref{eq:rhoDefinition}, \eqref{eq:EdHAppendix} yields
\begin{equation}\label{eq:EdH_L1}
\mathbb{E}[\Delta H] = \frac{\left( B^{\bmz}_h + C^{\bmz}_h \right)^2}{2}.
\end{equation}
For the 1-stage Velocity Verlet integrator \eqref{eq:1sVV}, one has
\begin{equation*}
\Psi_h^{\text{VV}} \left(\begin{array}{c}
\theta_{i} \\ p_{i}
 \end{array}\right) =
\left(\begin{array}{c}
\left( 1 - \frac{h^2}{2} \right) \theta_i + h p_i \\ \left( - h + \frac{h^3}{4} \right) \theta_i + \left(1 - \frac{h^2}{2} \right) p_i \end{array}\right),
\end{equation*}
that is
\begin{equation*}
B^{\bmz}_h = h, \qquad C^{\bmz}_h = - h + \frac{h^3}{4},
\end{equation*}
which, combined with \eqref{eq:EdH_L1}, provides
\begin{equation}\label{eq:EdH1sVV}
\mathbb{E}^1_{\text{VV}} [\Delta H] = \frac{h_{\text{VV}}^6}{32}.
\end{equation}

\section{SIR model}
\label{app:SIR}
The system of ODEs underlying the Susceptible-Infectious-Removed (SIR) compartmental model is the following
\begin{equation}\label{eq:SIR_model}
\begin{dcases}
\frac{dS}{dt} = - \beta S \frac{I}{P}, \\
\frac{dI}{dt} = \beta S \frac{I}{P} - \gamma I, \\
\frac{dR}{dt} = \gamma I,
\end{dcases}
\end{equation}
with initial conditions $S(t_0) = P - I_0$, $I(t_0) = I_0$ and $R(t_0) = 0$. Here, $S(t)$ is the number of the susceptibles, $I(t)$ is the number of infectious people, $R(t)$ is the number of recovered individuals, $\beta$ is the transmission rate, $\gamma$ is the inverse of the average infectious time, $I_0$ is the initial number of infectious individuals and $P = S(t) + I(t) + R(t)$ is the total (constant) population. Since we utilized daily incidence data gathered in the Basque Country during the COVID-19 pandemic, we added a counting compartment $C_I(t)$ which counts the number of new infections, i.e.
\begin{equation*}
\frac{dC_I}{dt} = \beta S \frac{I}{P}.
\end{equation*}
Due to the imprecise collection of data during the COVID-19 pandemic, we took explicitly into account under-reporting of new infected cases. Given $\{\tilde{C}_{t_0 + j}\}_{j=1}^n$ that account for the new daily incidence - $n$ is the number of days, in our case $n = 356$ - one can see them as realizations of a random variable $\tilde{C}(t_0 + j)$ which gives the daily incidence at day $j$. In that way, $\frac{\tilde{C}(t_0 + j)}{\eta (t_0 + j)}$, $j = 1, ..., n$ represents the real number of new daily infections, $\eta (t) \in (0, 1]$. Therefore, following \cite{inouzhe2023}, we took
\begin{equation*}
\frac{\tilde{C}(t)}{\eta (t)} \sim \text{NB} (C(t), \phi),
\end{equation*}
where $C(t) = C_I (t) - C_I (t-1)$ and $\phi^{-1}$ controls the overdispersion around $C(t)$.

For our numerical experiments, the model parameters to be estimated have the following priors:
\begin{equation*}
\beta \sim \mathcal{N} (\beta_{\mu}, \beta_{\sigma}), \quad \gamma \sim \mathcal{N} (\gamma_{\mu}, \gamma_{\sigma}), \quad I_0 \sim \mathcal{N} (I_{0_{\mu}}, I_{0_{\sigma}}), \quad \phi^{-1} \sim \text{Exp} (\phi^{-1}_{\lambda}),
\end{equation*}
where
\begin{align*}
& \beta_{\mu} = 0.3, \quad \beta_{\sigma} = 0.1, \\
& \gamma_{\mu} = 0.1, \quad \gamma_{\sigma} = 0.015, \\
& I_{0_{\mu}} = 21.88017, \quad I_{0_{\sigma}} = 7.29339, \\
& \phi^{-1}_{\lambda} = 0.1.
\end{align*}

The ODE system \eqref{eq:SIR_model} was solved numerically using CVODES from the SUNDIALS suite \cite{cvode_documentation}, which employs the backward differentiation formula (BDF) method, Newton iteration with the DENSE linear solver, and a user-supplied Jacobian routine (for details see \cite{cvode_documentation} and references therein).

\section*{Acknowledgments}
\begin{sloppypar}
We thank Tijana Radivojevi{\'c}, Jorge P{\'e}rez Heredia and Felix M{\"u}ller for their valuable contributions at the early stage of the study.
We thank Hristo Inouzhe and María Xosé Rodríguez-Álvarez for providing access to the data and for their contributions to the implementation of the SIR model in \textsf{HaiCS}.
We thank Martín Parga-Pazos for the discussions about the different approaches for the Effective Sample Size estimation.

We acknowledge the financial support by the Ministerio de Ciencia e Innovación, Agencia Estatal de Investigación (MICINN, AEI) of the Spanish Government through BCAM Severo Ochoa accreditation CEX2021-001142-S (LN, EA) and grants PID2019-104927GB-C22, PID2019-104927GB-C21, MCIN/AEI/10.13039/501100011033, ERDF (``A way of making Europe'') (JMSS). This work was supported by the BERC 2022-2025 Program (LN, EA), Convenio IKUR 21-HPC-IA (EA) and grants KK-2022/00006 (EA), KK-2021/00022 (EA, LN) and KK-2021/00064 (EA), and by La Caixa - INPhINIT 2020 Fellowship, grant LCF/BQ/DI20/11780022 (LN), funded by the Fundación 'la Caixa'. This work has been possible thanks to the support of the computing infrastructure of the i2BASQUE academic network, Barcelona Supercomputing Center (RES), DIPC Computer Center, BCAM in-house cluster Hipatia and the technical and human support provided by IZO-SGI SGIker of UPV/EHU.
\end{sloppypar}

\section*{References}
\bibliographystyle{elsarticle-num}
\bibliography{sAIApaper_bibliography}

\end{document}